\pdfoutput=1
\documentclass[a4paper,reprint,twocolumn,notitlepage,aip,cha,nofootinbib]{revtex4-2}
\usepackage[left=1.5cm,right=1.5cm,top=1cm,bottom=1.5cm,includeheadfoot]{geometry}
\usepackage[english]{babel}
\usepackage[T1]{fontenc}
\usepackage{tgtermes} 
\usepackage{tgheros} 

\usepackage{amsmath}
\usepackage{amssymb}
\usepackage{amsthm}
\usepackage{mathtools}
\usepackage{bm}

\usepackage{graphicx}
\usepackage[dvipsnames]{xcolor}
\usepackage{tikz}
\usepackage{pgfplots} 

\usepackage{multirow}
\usepackage{soul}
\usepackage{cancel}
\usepackage{framed}

\usepackage{tcolorbox}
\tcbset{
  simple/.style={
    colback=white,      
    colframe=black,     
    boxrule=0.5pt,      
    arc=0pt,            
    left=4pt,right=4pt, 
    top=4pt,bottom=4pt,
  }
}

\PassOptionsToPackage{hyphens}{url} 
\usepackage[breaklinks=true]{hyperref}
\bibpunct{[}{]}{;}{n}{}{} 

\usepackage{enumerate}
\usepackage[shortlabels]{enumitem}

\theoremstyle{remark}

\newtheorem{assumptions}{Assumption}

\theoremstyle{definition}

\newcommand{\rmi}{\mathrm{i}}
\newcommand{\rms}{\mathrm{s}}
\newcommand{\rmd}{\mathrm{d}}

\newcommand{\R}{\mathbb{R}}

\newcommand{\rr}{\bm{r}}
\newcommand{\kk}{\bm{k}}
\newcommand{\GG}{\bm{G}}
\newcommand{\eps}{\varepsilon}
\DeclareMathOperator{\trace}{Tr}
\newcommand{\prox}{\mathop\mathrm{prox}}

\newcommand{\argmin}{\mathop\mathrm{arg\,min}}
\newcommand{\id}{\mathop\mathrm{id}}
\newcommand{\weakto}{\rightharpoonup}

\newcommand{\FDM}{F_\mathrm{DM}}

\newcommand{\Exc}{E_\mathrm{xc}}
\newcommand{\Exceps}{E_{\mathrm{xc},\eps}}

\newcommand{\EH}{E_{\rm H}}
\newcommand{\EHartree}{E^{\rm Hartree}}
\newcommand{\FHartree}{F^{\rm Hartree}}
\newcommand{\Efield}{E^\mathrm{f}}
\newcommand{\EM}{E_{\rm M}}
\newcommand{\bEM}{{\bar E}_{\rm M}}
\newcommand{\bFM}{{\bar F}_{\rm M}}
\newcommand{\beM}{{\bar e}_{\rm M}}
\newcommand{\bfM}{{\bar f}_{\rm M}}

\newcommand{\vo}{v_\mathrm{off}}
\newcommand{\vxc}{v_\mathrm{xc}}

\newcommand{\vH}{v_\mathrm{H}}

\newcommand{\vxceps}{v_{\mathrm{xc},\eps}}
\newcommand{\norm}[1]{\left\| #1 \right\|}
\newcommand{\abs}[1]{\left| #1 \right|}
\renewcommand{\AA}{\bm{A}}
\renewcommand{\aa}{\bm{a}}

\newcommand{\bb}{\bm{b}}
\newcommand{\aalpha}{\bm{\alpha}}
\newcommand{\jj}{\bm{j}}
\newcommand{\jp}{\bm{j}^\mathrm{p}}
\newcommand{\rhogs}{\rho_\mathrm{gs}}
\newcommand{\rhogseps}{\rho_\mathrm{gs,\eps}}
\newcommand{\Fguide}{\mathcal{F}}
\newcommand{\tvxceps}{{\tilde v}_{\mathrm{xc},\eps}}

\begin{document}

\title{Perspective on Moreau--Yosida Regularization in\\Density-Functional Theory} 

\author{Markus Penz}
\affiliation{Arnold Sommerfeld Center for Theoretical Physics,
\mbox{Ludwig-Maximilians-Universität München, Germany}}
\affiliation{Munich Center for Quantum Science and Technology, Munich, Germany}
\affiliation{Department of Computer Science, Oslo Metropolitan University, Norway}
\affiliation{Max Planck Institute for the Structure and Dynamics of Matter, Hamburg, Germany}

\author{Michael F. Herbst}
\affiliation{Mathematics for Materials Modelling (MatMat), Institute of Mathematics \& Institute of Materials,
\'Ecole Polytechnique Fédérale de Lausanne, Switzerland}
\affiliation{National Centre for Computational Design and Discovery of Novel Materials (MARVEL),
\'Ecole Polytechnique F\'ed\'erale de Lausanne, Switzerland}

\author{Trygve Helgaker}
\affiliation{Hylleraas Centre for Quantum Molecular Sciences, Department of Chemistry, \mbox{University of Oslo, Norway}}

\author{Andre Laestadius}
\email[Electronic address:\;]{andre.laestadius@oslomet.no}
\affiliation{Department of Computer Science, Oslo Metropolitan University, Norway}
\affiliation{Hylleraas Centre for Quantum Molecular Sciences, Department of Chemistry, \mbox{University of Oslo, Norway}}

\begin{abstract}
Within density-functional theory, Moreau--Yosida regularization enables both a reformulation of the theory and a mathematically well-defined definition of the Kohn--Sham approach. It is further employed in density--potential inversion schemes and, through the choice of topology for the density and potential space, can be directly linked to classical field theories. This perspective collects various appearances of the regularization technique within density-functional theory alongside possibilities for their future development.
\end{abstract}

\maketitle
\tableofcontents

\section{Introduction}
\label{sec:intro}

It has been more than 10 years since Moreau--Yosida (MY) regularization was introduced into density-functional theory (DFT)~\cite{Kvaal2014}, and it has arguably turned out to be an unexpectedly fruitful extension. First introduced as a conceptual tool to fix differentiability issues of the exact functional and to make the exchange--correlation potential mathematically well defined, it was later realized that it can be seen as the theoretical basis of density--potential inversion techniques. In order to connect it to existing methods, like the Zhao--Morrison--Parr inversion scheme~\cite{Zhao1994}, the topology of the density and potential spaces needed to be adjusted in a way that is critically different from the standard setting of DFT introduced by \citet{Lieb1983} based on Lebesgue spaces $L^p$. This recent realization~\cite{Penz2023-MY-ZMP} marked an important shift of perspective, since it became apparent that the topology of those spaces can entail physical meaning. More concretely, the duality map that canonically connects the density space and the potential space as its dual, is just the Poisson equation of electrostatics if the spaces are chosen as special Sobolev spaces. In this way, new physical meaning and connections to classical fields enter the theory and need to be interpreted accordingly. At this stage of development, many exciting routes can be taken, be it about the precise choice of topology, the extension of the spaces to include vector potentials, or the utilization of the regularized theory for computational purposes. 

This perspective article aims at showing the current state of the field as well as mapping out several directions for continued research that we deem promising. We believe that, as the theory matures, it can establish itself as a mathematically rigorous but also computationally useful extension of DFT. Ideally, this means offering new and enhanced approximations for electronic-structure calculations, while also covering a wider range of physical phenomena. A main result of this work (discussed in Section~\ref{sec:formalDFT}) is that the necessary function-space framework for MY regularization can be combined with the usual potential space of quantum mechanics only if the setting is such that it always guarantees existence of a ground state. Another interesting finding, laid out in Section~\ref{sec:MY-mixing}, is the possible connection of the topology to physics, in that the norm of the density and potential spaces exactly gives their energy content and that they are connected via the Poisson equation. These and many more insights into MY regularization applied to DFT are discussed in this perspective article, where we also make the relevant connections to the literature.

Since the original intention of introducing MY regularization into DFT was to make the theory mathematically sound, it is necessary to introduce a certain amount of mathematical concepts. We try to present this in the preliminary Section~\ref{sec:pre} with minimal requirements, but the interested reader might want to follow up on certain details by consulting the vast mathematical literature on the fields that we mainly rely on. These are functional and convex analysis formulated on Banach and Hilbert spaces, as well as the more specialized theory of Sobolev spaces, and the reference list includes our favorite textbooks~\cite{Barbu-Precupanu,Bauschke-Combettes,zalinescu2002,rockafellar2009variational,chidume-book,Blanchard-Bruening,adams-book,tiel1984}. After covering the most important mathematical topics in Sections~\ref{sec:func-anal}, \ref{sec:conv-anal}, and \ref{sec:MY}, we continue in Section~\ref{sec:formalDFT} with a formalized presentation of DFT that introduces the universal density functional for a fully interacting system and for an auxiliary, non-interacting system. This is complemented by Section~\ref{sec:ks} on the Kohn--Sham (KS) method, which we introduce with an additional offset potential to capture \emph{a priori} information such as external binding potentials.
Section~\ref{sec:MY-DFT} is concerned with introducing MY regularization into DFT in three different ways---namely, based on the universal density functional, based on the total energy functional, and based on density--potential mixing. The next two sections describe the main use cases for MY regularization in DFT to date: a technique for density--potential inversion in Section~\ref{sec:inversion} and a rigorous formulation of the KS method that takes a central role in the application of DFT in Section~\ref{sec:regKS}. Section~\ref{sec:autoreg} shows how certain formulations of DFT, such as Hartree theory and Maxwell--Schrödinger DFT, get automatically regularized through the inclusion of a mean-field energy contribution. Section~\ref{sec:MY-periodic} deals with the numerical realization of density--potential inversion in the KS setting for periodic systems. Finally, we conclude with an extended outlook in Section~\ref{sec:outlook}.
We provide a brief notational aid for the most frequently used symbols in Table~\ref{tab:notation}.

\begin{table}[h!]
\begin{tcolorbox}[simple]
\caption{\textbf{Notational aid for frequently used symbols.}\\\vspace{.5em}}
\begin{tabular}{llll}
     $X$ & generic Banach space for densities & \multirow{2}{*}{$\begin{drcases*} \\ \\ \end{drcases*}$} & Sec.~\ref{sec:func-anal} \&\\
     $X^*$ & generic Banach space for potentials && Sec.~\ref{sec:formalDFT} \\
     $D$ & set of $N$-representable densities && Eq.~\eqref{eq:D-def}\\
     $J$ & duality map from $X$ to $X^*$ && Eq.~\eqref{eq:J-def}\\
     $f^*(u)$ & Legendre--Fenchel transform && Eq.~\eqref{eq:def-LF} \\
     & (convex conjugate) \\
     $\partial f(x)$ & sub/superdifferential of a && Eq.~\eqref{eq:subdiff} \\
     & convex/concave functional \\
     $f_\eps(x)$ & MY regularization && Eq.~\eqref{eq:MY}\\
     $\nabla f_\eps(x)$ & functional derivative && Sec.~\ref{sec:MY}\\
     $\prox_{\eps f}(x)$ & proximal map && Eq.~\eqref{eq:def-prox}\\
     $F^\lambda(\rho)$ & universal functional && Eq.~\eqref{eq:DFT-func-2} \\
     $E^\lambda(v)$ & ground-state energy functional && Eq.~\eqref{eq:E-def} \\
     $F_\eps^\lambda(x)$ & MY-regularized universal functional && Eq.~\eqref{eq:Feps} \\
     $E_\eps^\lambda(v)$ & convex conjugate of $F_\eps^\lambda(x)$ && Eq.~\eqref{eq:Eeps-def} \\
\end{tabular}
\end{tcolorbox}
\label{tab:notation}
\end{table}

\section{Preliminaries}
\label{sec:pre}

The following five subsections introduce the mathematical and theoretical prerequisites for performing MY regularization in DFT and for studying related topics. Section~\ref{sec:func-anal} deals with some basic notions of functional analysis, most importantly certain properties of Banach spaces that are later required for MY regularization as well as defining the duality map. Section~\ref{sec:conv-anal} gives some central results needed for the convex analysis of DFT. Section~\ref{sec:MY} is the last mathematical part and introduces MY regularization with its most important properties.
Section~\ref{sec:formalDFT} continues with a presentation of formal DFT and contains an important result that connects the properties of the function spaces with the existence of ground states. Finally, Section~\ref{sec:ks} briefly introduces the KS method of DFT.

\subsection{Functional analysis preliminaries}
\label{sec:func-anal}

Our development of MY regularization applied to DFT relies heavily on topological properties of the involved density and potential spaces, the Banach space $X$ and its topological dual $X^*$. We keep them as abstract spaces for most of our presentation, since this allows the application of the results to a huge variety of different settings. It will be indispensable to repeat important results from functional analysis, especially from the theory of Banach spaces, since this will be the first pillar on which we base DFT mathematically. The properties of the spaces $X$ and $X^*$ are collected in three different assumption sets of increasing restrictiveness, to which later results will refer.

The basic setting for the whole presentation is a real Banach space $X$ with the norm denoted as $\|\cdot\|_X$. Banach spaces have been developed from the beginning as spaces of functions~\cite{Banach1922} and as such typically are infinite dimensional, which complicates their mathematical study. The feature that makes a normed vector space a Banach space is that it is \emph{complete}, meaning that every Cauchy sequence in $X$ has a well-defined limit within the space $X$.
Convergence to a limit value usually means \emph{strong convergence}---that is, convergence in the norm $\|\cdot\|_{X}$, where $x_n \to x$ in $X$ if and only if
\begin{equation}
    \lim_{n\to\infty}\|x_n-x\|_X=0.
\end{equation}
Since the norm of the space naturally induces a topology (a notion of `closeness' and `continuity'), we are able to define the \emph{topological dual} of $X$ as the space of linear and continuous maps from $X$ to $\R$ (called `functionals'),
\begin{equation}
    X^* = \{u\colon X\to\R\mid u\;\text{linear and continuous}\}.
\end{equation}
Now, $X^*$ gets equipped with a norm itself, the operator norm on $X$, which we will accordingly denote with $\|\cdot\|_{X^*}$,
\begin{equation}
    \|u\|_{X^*} = \sup_{\|x\|_X\leq 1}|u(x)|.
\end{equation}
The completeness of $X$ then carries over to $X^*$, making it a Banach space as well. With the help of $X^*$, we can define \emph{weak convergence} $x_n \weakto x$ in $X$ as the property that, for all $u\in X^*$, it holds that
\begin{equation}
    \lim_{n\to\infty} u(x_n-x) = \lim_{n\to\infty} u(x_n) - u(x) = 0.
\end{equation}
As suggested by the terminology, strong convergence of a sequence implies weak convergence.
Further, let
\begin{equation}
    \label{eq:J-def}
    J(x) = \{u\in X^* \mid u(x)=\|x\|_{X}^2=\|u\|_{X^*}^2\}
\end{equation}
be a set-valued map that maps each element in $X$ to a subset of $X^*$, called the \emph{duality map}. This generalizes the Riesz representation for Hilbert spaces, that is a one-to-one map, to the case of Banach spaces. The Hahn--Banach theorem guarantees that the image of every $x\in X$ under $J$ is non-empty and it is easy to see that $J$ is  homogeneous (i.e., $J(\alpha x)=\alpha J(x)$ for every $\alpha\in\R$), but we would like to impose further topological properties on $X$ so as to make the duality mapping nicer. A Banach space for which $J(x)$ is always single-valued (and thus has \emph{not} to be considered as a set-valued map) is called \emph{smooth}. 
In a smooth Banach space, the duality map is  unique and given by the Gâteaux (functional) derivative of the convex function $\phi(x)=\frac{1}{2}\|x\|_{X}^2$,
\begin{equation}\label{eq:J-from-subdiff}
    J(x)=\nabla\phi(x).
\end{equation}
Interestingly, the smoothness of a space $X$ implies that its dual $X^*$ is \emph{strictly convex} (meaning that the line segment between any two points $u\neq v$, $\|u\|_{X^*}=\|v\|_{X^*}=1$,
lies strictly within the unit ball). This is an expression of the general connection of smoothness properties of a Banach space to convexity features on the dual side.

The next important property for our space $X$ is that of \emph{reflexivity}, which means that the bidual $X^{**}=(X^*)^*$ is isometrically isomorphic to the space $X$ itself and can be identified with it. Hence, if $X$ is reflexive, then $X$ can be recovered from $X^\ast$ as $X = X^{**} = (X^*)^*$.

If $X$ is reflexive, then $X^*$ is also reflexive.
We may therefore also set up a duality map $J^\ast$ from $X^\ast$ to $X= X^{\ast\ast}$. For every $u\in X^*$, we can consider an $x\in J^*(u) = \{x\in X\mid u(x)=\|x\|_{X}^2=\|u\|_{X^*}^2\}$ and with this recover $u\in J(x)$, so $J$ is also surjective if $X$ is reflexive. For a surjective function, we can consider its set-valued inverse and we thus get $J^{-1}=J^*$.
Now note that, when $X^*$ is also smooth ($X$ strictly convex), we have $J^{-1}$ single-valued as well, so $J$ is indeed bijective~\cite[Prop.~1.117]{Barbu-Precupanu}.
We collect all these properties as the minimal assumptions for our Banach spaces $X$ and $X^*$:

\begin{tcolorbox}[simple]
\begin{assumptions}\label{ass:BS-reflexive}
    The real Banach space $X$ is reflexive, with $X$ and $X^*$ strictly convex. The corresponding duality map $J\colon X\to X^*$ is then single-valued with a well-defined inverse $J^{-1}\colon X^*\to X$.
\end{assumptions}
\end{tcolorbox}

The prototypical example for such Banach spaces, which is also relevant in the context of DFT, are the Lebesgue spaces $L^p$, $1<p<\infty$. It is important to note that $p=1$ and $p=\infty$ are excluded by Assumption~\ref{ass:BS-reflexive}, since those spaces are neither reflexive nor strictly convex. At this point, it already becomes obvious that we need to adapt the typical choice of density space in DFT~\cite{Lieb1983}, $L^1(\R^3)\cap L^3(\R^3)$, since by including $L^1$ the space becomes non-reflexive and thus cannot fulfill Assumption~\ref{ass:BS-reflexive}. It is precisely for this reason that, for example, \citet{eschrig2010t} in his discussion of thermal DFT switches to the reflexive space $L^3(\Omega)$ on a bounded three-dimensional periodic domain, which anyway entails $L^3(\Omega)\subset L^1(\Omega)$.

Assumption \ref{ass:BS-reflexive} has been our standard assumption in several previous publications on Moreau--Yosida regularization in DFT~\cite{KSpaper2018,CDFT-paper,Penz2023-MY-ZMP} and is sufficient for many use cases. Yet, in some instances, additional regularity is required and we add the  condition that $X^*$ is also uniformly convex. Note that uniform convexity implies reflexivity. A space $X^*$ is  called uniformly convex if, for each $\eps\in (0,2]$, there is a $\delta(\eps)>0$ such that, for all $u,v\in X^*$ with $\|u\|_{X^*}=\|v\|_{X^*}=1$ and $\|u-v\|_{X^*}=\eps$, we have $1-\frac{1}{2}\|u+v\|_{X^*}\geq \delta(\eps)$. Adding the requirement of uniform convexity, we obtain:

\begin{tcolorbox}[simple]
\begin{assumptions}\label{ass:BS-reflexive-unif-convex}
    $X$ and $X^*$ fulfill all requirements of Assumption~\ref{ass:BS-reflexive} and additionally $X^*$ is assumed uniformly convex. The duality map $J\colon X\to X^*$ is then additionally uniformly continuous.
\end{assumptions}
\end{tcolorbox}

Finally, we can tighten those assumptions even more and move to a Hilbert space $X$, equipped with an inner product denoted as $\langle\cdot,\cdot\rangle_X$. Hilbert spaces are automatically uniformly convex and thus also reflexive. Since then also $X^*$ is a Hilbert space, we have all the properties in Assumption~\ref{ass:BS-reflexive-unif-convex}  fulfilled. For Hilbert spaces, the duality map $J\colon X\to X^*$ is exactly the isometric isomorphism from the Riesz representation theorem and linear as such. Nevertheless, we do not want to identify a Hilbert space with its dual in the given context.
The original work introducing Moreau--Yosida regularization to DFT of \citet{Kvaal2014} considered a Hilbert-space setting and it is also employed elsewhere~\cite{Cances2012,Sutter2024,CarvalhoCorso2025,Herbst2025,Bohle2026}. We also highlight a comprehensive chapter by \citet{kvaal2022-chapter} in a recent textbook on DFT.

\begin{tcolorbox}[simple]
\begin{assumptions}\label{ass:HS}
    The Banach space $X$ and consequently also $X^*$ are Hilbert spaces. The  duality map $J\colon X\to X^*$ is then linear and Lipschitz continuous with constant 1, with a well-defined inverse $J^{-1}\colon X^*\to X$ that has the same properties.
\end{assumptions}
\end{tcolorbox}

For each $u\in X^*$, the element $x=J^{-1}(u)\in X$ is then just the unique element given from the Riesz representation theorem and we have for every $y\in X$ that
\begin{equation}
    \label{eq:Riesz-rep}
    u(y) = \langle u,y \rangle = \langle x,y \rangle_{X}.
\end{equation}
Note carefully that the notation $\langle u,y \rangle$, which we will also employ in the Banach-space case from here on, describes the dual pairing of $u\in X^*$ with $y\in X$ (i.e., the application of the linear functional $u$ to $y$), and should not be confused with $\langle x,y \rangle_{X}$, which is the inner product of elements in the Hilbert space $X$.
This notation fits well with the previously mentioned case $X=L^p(\Omega)$ on some domain $\Omega\subseteq\R^n$, $1<p<\infty$, where $X^*=L^{p^*}(\Omega)$ with $1/p+1/p^*=1$, and 
\begin{equation}
    \label{eq:L2-pairing}
    u(x) = \langle u,x \rangle = \int_\Omega u(\rr)x(\rr)\rmd\rr,
\end{equation}
which equals $\langle u,x \rangle_{L^2}$ for $p=2$.
So, in this case, we arrive back at the inner product on the Hilbert space $L^2(\Omega)$, which is called a \emph{pivot space} between $L^p(\Omega)$ and $L^{p^*}(\Omega)$. In other words, the inner product of the pivot space serves as the duality pairing. 

This construction can be generalized to arbitrary $X$ and $X^*$ within the formalism of a \emph{rigged Hilbert space (Gel'fand triple)}. Applied to the Hilbert space of wavefunctions, this approach provides a rigorous foundation for Dirac's bra--ket formalism and can be employed to describe continuous spectra and scattering states in quantum mechanics~\cite{bohm1978rigged,Madrid2005}. Those are different objects than  the densities and potentials considered here, but we find the similarities still striking.

Note that the three assumption sets are increasingly strong---that is,  Assumption~\ref{ass:HS} implies Assumption~\ref{ass:BS-reflexive-unif-convex}, which in turn implies our minimal requirement Assumption~\ref{ass:BS-reflexive}.

\subsection{Convex analysis preliminaries}
\label{sec:conv-anal}

The second mathematical pillar of DFT is without any doubt convex analysis, since DFT admits a beautiful formulation in terms of convex and concave functionals on Banach spaces.
On a Banach space $X$ that fulfills Assumption~\ref{ass:BS-reflexive}, we now consider convex functions $f\colon X\to\R\cup\{+\infty\}$ that are not $+\infty$ everywhere. We also usually assume lower semicontinuity, which means that, for each $x_0\in X$, it holds that $\liminf_{x\to x_0} f(x) \geq f(x_0)$. Since the argument $x\in X$ of $f(x)$ is the element of a Banach space of functions that gets mapped to the scalar field, $f$ is sometimes called a \emph{functional} (especially in the context of DFT).
The \emph{Legendre--Fenchel transform} (convex conjugate) of any $f\colon X\to\R\cup\{-\infty,+\infty\}$ is a function $f^*\colon X^*\to\R\cup\{-\infty,+\infty\}$ given by
\begin{equation}\label{eq:def-LF}
    f^*(u) = \sup_{x\in X}\{\langle u,x \rangle - f(x)\}.
\end{equation}
The convex conjugate of a convex and lower semicontinuous $f\colon X\to\R\cup\{+\infty\}$ is again convex and lower semicontinuous.
By applying this transformation twice and restricting the result again to $X\subseteq X^{**}$ (biconjugate),
\begin{equation}
    f^{**}(x) = \sup_{u\in X^*}\{\langle u,x \rangle - f^*(u)\},
\end{equation}
one arrives back at the original function, $f^{**}=f$, if and only if $f$ is convex and lower semicontinuous.

The \emph{subdifferential} of $f$ at some $x\in X$ is the set-valued map
\begin{equation}\label{eq:subdiff}
    \partial f(x) = \{u\in X^*\mid \forall y \in X\colon f(y)\geq f(x) + \langle u,y-x \rangle\}.
\end{equation}
On the real line, the set $\partial f(x)$ can be interpreted as the collection of all slopes for which the corresponding tangent of $f$ at $x$ never exceeds the function $f$ itself (globally, since the function is assumed convex).
It can be defined for concave functions as the \emph{superdifferential} by just reverting the inequality, and we use the same notation in this case.
Note that $\partial f(x)$ is always a convex set, and that it includes infinitely many elements where $f$ has a kink and is thus non-differentiable. It can also be the empty set. To illustrate this, take $X=\R$ and $f(x)=-\sqrt{x}$ for $x\geq 0$ and $f(x)=+\infty$ for $x<0$ (which actually connects to reduced-density-matrix functional theory~\cite{Schilling2019}) and then note that $\partial f(0) = \emptyset$. Another one-dimensional example can be taken from a DFT-treatment of the quantum Rabi model~\cite{Bakkestuen2025}, where $\partial f(\pm 1)=\emptyset$ for $f(x)=-\sqrt{1-x^2}$ for $x\in[-1,1]$ and $f(x)=+\infty$ outside.
If, on the other hand, $f$ is differentiable at $x$, then the subdifferential $\partial f(x)$ is single-valued and consists only of the functional derivative $\nabla f(x)$. An graphical illustration with a function on $\R$ is provided in Figure~\ref{fig:convex-example}.

The subdifferential fulfills the usual role of a derivative, by qualifying the minimum of a convex (not necessarily lower semicontinuous) functional,
\begin{equation}
    0 \in \partial f(x_*) \iff x_* \in \argmin_{x\in X}f(x).
\end{equation}
Yet, unlike the derivative, the subdifferential does not require the function to be differentiable. For a function $g$ on the dual space $X^\ast$, we define the subdifferential at $u \in X^\ast$ as a convex subset of the predual $X$ rather than the bidual $X^{\ast\ast}$,
\begin{equation}\label{eq:subdiff-dual}
    \partial g(u) = \{x\in X\mid \forall v \in X^*\colon g(v)\geq g(u) + \langle v-u,x \rangle\}.
\end{equation}
With these definitions in place, and if $f$ is lower semicontinuous, the subdifferentials of a pair of conjugate functions $(f,f^*)$ are connected in the following way (inverse subdifferential relation),
\begin{equation}\label{eq:inv-subdiff}
    (\partial f)^{-1} = \partial f^*,
\end{equation}
where the inverse is meant as a set-valued mapping.

\subsection{Moreau--Yosida regularization}
\label{sec:MY}

In order to deal with problems arising from a possible non-differentiability of $f\colon X\to\R\cup\{+\infty\}$ , we introduce the MY regularization with smoothing parameter $\eps>0$ of a convex and lower semicontinuous functional $f$ as 
\begin{equation}\label{eq:MY}
    f_\eps (x) = \inf_{y\in X}\left\{ f(y) + \frac{1}{2\eps}\|x-y\|_{X}^2 \right\}.
\end{equation}
\begin{figure}[t]
\centering
\usetikzlibrary{decorations.pathreplacing}
\begin{tikzpicture}[scale=1.1]
    \begin{axis}[restrict y to domain=-.2:3,xlabel={},ylabel={},xticklabels=\empty,yticklabels=\empty,legend cell align=left,xlabel={$x$},xlabel near ticks,xlabel shift=-5pt]
    \addplot[no markers,color=MidnightBlue,thick] table[x index=0,y index=1] {MY-example-data2.dat};
    \addplot[no markers,color=MidnightBlue,thick,dashed] table[x index=0,y index=2] {MY-example-data2.dat};
    \addplot[no markers,color=red,thick] table[x index=0,y index=3] {MY-example-data2.dat};
    \addplot[no markers,color=MidnightBlue!50] table[x index=0,y index=4] {MY-example-data2.dat};
    \addplot[no markers,color=MidnightBlue!50] table[x index=0,y index=5] {MY-example-data2.dat};
    \addplot[no markers,color=MidnightBlue!50] table[x index=0,y index=6] {MY-example-data2.dat};
    \legend{$f(x)$,$f_\eps(x)$,$f(p)+\frac{1}{2\eps}\|x-p\|^2_X$}
    \end{axis}
    \draw[->] (1.38,2) node[below] {$x_0$} -- (1.38,2.4);
    \draw[->] (1.78,1) node[below] {$p=\prox(x_0)$} -- (1.78,2.2);
    \draw[->] (4.87,0.8) node[below] {$x_1$} -- (4.87,1.22);
    \draw [decorate,decoration={brace,amplitude=5pt}]
        (5.76,3.65) -- (5.76,1.72)
        node[midway,xshift=20pt]{$\partial f(x_1)$};
\end{tikzpicture}
\caption{Illustration of MY regularization of a convex $f(x)$ together with one of the parabolas that trace out $f_\eps(x)$. The parabola touches $f_\eps(x)$ at $x_0$ and has its vertex at the proximal point $\prox(x_0)$. Further, three elements from the subdifferential at $x_1$ are depicted that are linear functions that fully lie below the convex functional.}
\label{fig:convex-example}
\end{figure}
MY regularization was originally introduced by \citet{moreau1965} and can be linked to the previously discovered Yosida approximation of maximal monotone operators~\cite{yosida1948}.
How such a MY regularized functional looks like is shown with one-dimensional examples in Figure~\ref{fig:convex-example} and Figure~\ref{fig:MY-example} in Section~\ref{sec:MY-univ}.
This construction has a long list of favorable properties that we reproduce here, while the proofs can be found in standard textbooks~\cite{Barbu-Precupanu,zalinescu2002}. It is important to remember that we always demand at least Assumption~\ref{ass:BS-reflexive}, ensuring that $X$ and $X^\ast$ are both reflexive and strictly convex:
\begin{enumerate}[(i)]
    \item\label{item:diff} $f_\eps$ is convex, finite, and Gâteaux differentiable. If in addition $X^*$ is uniformly convex (Assumption~\ref{ass:BS-reflexive-unif-convex}), then $f_\eps$ is also Fréchet differentiable.
    \item\label{item:inf} For each $x\in X$ and all $\eps,\eps^\prime$ such that $0 < \eps' < \eps$, we have $\inf f(X) \leq f_\eps(x) \leq f_{\eps'}(x) \leq f(x)$ . In particular, $\inf f(X)=\inf f_\eps(X)=\inf f_{\eps'}(X)$. The minimal value and the minimizers, if they exist, are preserved by the regularization.
    \item\label{item:limit} $\lim_{\eps\to 0} f_\eps(x)=f(x)$ for every $x\in X$.
    \item The infimum in Eq.~\ref{eq:MY} is  attained at a unique point (\emph{proximal point}) that defines the \emph{proximal map}
    \begin{equation}\label{eq:def-prox}
    \begin{aligned}
    & \qquad \prox_{\eps f} \colon  X \to X, \\
     &   \qquad \prox_{\eps f}(x) = \argmin_{y\in X}\left\{ f(y) + \frac{1}{2\eps}\|x-y\|_{X}^2 \right\},
    \end{aligned}
    \end{equation}
    and $\lim_{\eps\to 0} \prox_{\eps f}(x) = x$ if $f(x)<\infty$.
        \item\label{item:diff-f_eps} $\nabla f_\eps(x) = \eps^{-1} J(x-\prox_{\eps f}(x)) \in \partial f(\prox_{\eps f}(x))$.
        \item $f_\eps (x) = f(\prox_{\eps f}(x)) + \frac{\eps}{2}\|\nabla f_\eps(x)\|_{X^*}^2$. 
    \item\label{item:diff-f_eps-convergence} For each $x\in X$ with $\partial f(x)$ non-empty, it holds that $\lim_{\eps\to 0}\nabla f_\eps(x) = w$ (weakly) where $w\in X^*$ is the (unique) element in $\partial f(x)$ with minimal norm. If $X^*$ is uniformly convex (Assumption~\ref{ass:BS-reflexive-unif-convex}), then the convergence is strong.
    \item For the Legendre--Fenchel transform of $f_\eps$, it holds that
    \begin{align}
        &(f_\eps)^*(u) = f^*(u) + \tfrac{\eps}{2}\|u\|_{X^*}^2 \quad\text{and} \label{eq:feps-conj} \\
        &\partial (f_\eps)^*(u) = \partial f^*(u) + \eps J^{-1}(u) \label{eq:diff-feps-conj}
    \end{align}
    for each $u\in X^*$. By adding $\tfrac{\eps}{2}\|u\|_{X^*}^2$ to $f^*(u)$, the $(f_\eps)^*$ automatically becomes a strictly convex function.
\end{enumerate}

Note that $X$ strictly convex was already needed for the definition of the proximal map since it means that $x\mapsto\|x\|^2_X$ is strictly convex and thus the minimum in Eq.~\eqref{eq:def-prox} is reached at a unique point. The reason that we can write $\prox_{\eps f}$ instead of $\prox_{f}^{\eps}$ (or similar) in order to include the regularization parameter $\eps$ in the notation, is that the definition of the proximal map could have also been written as
\begin{equation}
    \prox_{\eps f}(x) = \argmin_{y\in X}\left\{ \eps f(y) + \frac{1}{2}\|x-y\|_{X}^2 \right\}.
\end{equation}

In a Hilbert-space setting (Assumption~\ref{ass:HS}), a few additional useful properties of the MY regularization and the proximal map can be added~\cite{Bauschke-Combettes}:

\begin{enumerate}[(i)]
    \setcounter{enumi}{8}
    \item\label{item:frechet} $f_\eps$ is Fréchet differentiable and its functional derivative $\nabla f_\eps$ is Lipschitz continuous with constant $\eps^{-1}$.
    \item\label{item:prox-resolvent} The proximal map is given by the resolvent of the sub{\-}differential, $\prox_{\eps f}(x) = (\id + \eps\partial f)^{-1}(x)$.
    \item\label{item:prox-nonexpansive} The proximal map is firmly nonexpansive, meaning that, for all $x,y\in X$, it holds
    \begin{equation}\label{eq:prox-nonexpansive}
    \begin{aligned}
        &\langle \prox_{\eps f}(x)-\prox_{\eps f}(y),x-y \rangle_X \\
        &\qquad\qquad \geq \|\prox_{\eps f}(x)-\prox_{\eps f}(y)\|^2_X.
    \end{aligned}
    \end{equation}
    The same property holds for $\id - \prox_{\eps f}$.
    \item\label{item:strong-mon}
    $\partial (f_\eps)^*$ is strongly monotonous, meaning that, for all $u,v\in X^*$ and $x\in\partial (f_\eps)^*(u)$, $y\in\partial (f_\eps)^*(v)$, it holds
    \begin{equation}\label{eq:str-mon}
        \langle u-v,x-y \rangle \geq \eps\|u-v\|^2_{X^*}.
    \end{equation}
\end{enumerate}

While the notion of firm nonexpansiveness itself can be generalized to Banach spaces~\cite{Kohsaka2008}, it does not necessarily hold for the proximal map anymore. Nevertheless, a result of uniform continuity can be achieved for the proximal map also in a certain Banach-space setting~\cite{bacak-kohlenbach-2018}. Combining \ref{item:diff-f_eps} from the previous properties and the resolvent form for the proximal map, we obtain
\begin{equation}
    \nabla f_\eps(x) = \eps^{-1}\left(x - (\id + \eps\partial f)^{-1}(x)\right),
\end{equation}
which exactly corresponds to the previously mentioned Yosida approximation~\cite[Def.~23.1]{Bauschke-Combettes} of the multivalued mapping $\partial f$. Together with Moreau's contributions to convex analysis this justifies the naming of the regularization method.

\subsection{Formal density-functional theory}
\label{sec:formalDFT}

The standard setting of DFT is the $N$-particle Hamiltonian with an external, scalar potential $v$ on $\R^3$ (atomic units):
\begin{equation} \label{eq:Ham}
\begin{aligned}
    &\hat H^\lambda_v = \hat T + \lambda \hat W + \sum_{j=1}^N v(\rr_j),\\
    &\hat T =-\frac 1 2 \sum_{j=1}^N \nabla_j^2, \quad \hat W =\sum_{j<k} \vert \rr_j-\rr_k\vert^{-1}.
\end{aligned}
\end{equation}
Here, we allow different choices of the interaction strength $\lambda$, which can range from the non-interacting case ($\lambda=0$) to the fully interacting case with a repulsive force ($\lambda = 1$). For all choices of $\lambda\in\R$ and $v\in L^{3/2}(\R^{3})+L^\infty(\R^3)$, this Hamiltonian is self-adjoint and bounded below on its usual form domain by the Kato--Lions--Lax-Milgram--Nelson (KLMN) theorem~\cite[Th.~X.17 and Th.~X.19]{reed-simon-2}.
It was already mentioned that the methods and results presented here readily apply to various settings outside of standard DFT, although subtleties around the involved spaces and definitions of the basic functionals have to be taken into account. A setting that also involves vector potentials is briefly discussed in Section~\ref{sec:mdft}.

The ground-state problem of quantum mechanics is given by the Rayleigh--Ritz variational principle,
\begin{equation}\label{eq:E-def}
    E^\lambda(v) = \inf_{\Psi} \langle \Psi, \hat H^\lambda_v \Psi\rangle = \inf_{\hat\Gamma}\trace(\hat H^\lambda_v\hat \Gamma),
\end{equation}
where the variation is over all normalized wave functions or density matrices with finite kinetic energy and the required antisymmetry. The result, $E^\lambda(v)$, is then the ground-state energy of the system for a given external potential $v$ and chosen interaction strength $\lambda$. The principal idea of DFT is now to separate off the only density-dependent part inside this variation, while the remaining part is \emph{universal} in the sense that it does not depend on $v$~\cite{Hohenberg1964,percus1978,levy1979,Lieb1983}. An elegant way to achieve this is by Levy's constrained search---that is, one varies over all wave functions $\Psi$ or density matrices that yield a given one-particle density that we define for wave functions and density matrices $\hat\Gamma=\sum_j p_j|\Psi_j\rangle\langle\Psi_j|$ as
\begin{align}
    \rho_\Psi(\rr) &= N\int|\Psi(\rr,\rr_2,\ldots,\rr_N)|^2\rmd\rr_2\ldots\rmd\rr_N, \\
    \rho_{\hat\Gamma}(\rr) &= \sum_j p_j \rho_{\Psi_j}(\rr) ,\quad p_j\in [0,1], \quad \sum_j p_j = 1.
\end{align}
We define the universal density functional by a constrained search over pure states and ensembles, respectively,
\begin{align}
    &\tilde F^\lambda(\rho) = \inf_{\Psi\mapsto\rho}\langle\Psi, \hat H^\lambda_0 \Psi \rangle, \label{eq:F-pure} \\
    &\FDM^\lambda(\rho) = \inf_{\hat\Gamma\mapsto\rho}\trace(\hat H^\lambda_0\hat \Gamma),
    \label{eq:F-ens}
\end{align}
where $\hat H^\lambda_0 = \hat T + \lambda \hat W$ and where we write $\Psi\mapsto\rho$ for an $N$-particle wave function with $\rho_\Psi = \rho$ and analogously for density matrices.
Since the constrained search over ensembles includes the search over pure states, we have $\FDM^\lambda(\rho)\leq \tilde F^\lambda(\rho)$.
The functional $\FDM^\lambda(\rho)$ is convex since it is given as the infimum of a linear functional over a convex set, unlike $\tilde F^\lambda(\rho)$, which is \emph{not} necessarily convex. At non-pure-state $v$-representable $\rho$ it holds $\tilde F^\lambda(\rho)\neq \FDM^\lambda(\rho)$~\cite{Levy1982,Lieb1983}, and explicit examples are known~\cite[Sec.~VI.E]{Penz2021-GraphDFT}.

The effective domain of a convex functional is the set of points where $f(x)<+\infty$.
For the constrained-search functionals $\tilde F^\lambda(x)$ and $\FDM^\lambda(x)$, the effective domains agree and are also independent of $\lambda$ and we set
\begin{align}\label{eq:D-def}
    D &= \{x\in X\mid \tilde F^\lambda(x)<\infty\} \nonumber \\&= \{x\in X\mid \FDM^\lambda(x)<\infty\}.
\end{align}
From the definition of the density $\rho_\Psi$ and the definition of the functionals, it follows that any $\rho\in D$ must be non-negative and normalized to the number of particles. It is a fact that $D$ corresponds to the \emph{$N$-representable} densities~\cite[Th.~1.2]{Lieb1983}.
Note that we generally use $\rho$ for elements in $D$ that are real densities, while we use $x$ for more general elements in $X$ that will later rise to prominence as `mixed densities'.
By definition, we have $\tilde F^\lambda(x)=\FDM^\lambda(x)=+\infty$ for $x\in X\setminus D$. For this reason, the usual convention will be to use the symbol $\rho$ for elements of $D$ that are `real' densities, and to use $x,y\in X$ otherwise.


The pure- and ensemble-state density functionals given in Eqs.~\eqref{eq:F-pure} and \eqref{eq:F-ens}, respectively, yield the same ground-state energy in the Hohenberg--Kohn variational principle,
\begin{equation}\label{eq:E-var}
\begin{aligned}
    E^\lambda(v) &= \inf_{\rho\in X} \{ \tilde F^\lambda(\rho) + \langle v,\rho \rangle \}   \\ &= \inf_{\rho\in X} \{ \FDM^\lambda(\rho) + \langle v,\rho \rangle \},
\end{aligned}
\end{equation}
where the dual pairing of potential and density yields the potential energy, $\langle v,\rho \rangle=\int v(\rr)\rho(\rr)\rmd\rr$.
Recognizing that Eq.~\eqref{eq:E-var} has the structure of the Legendre--Fenchel transform introduced in Section~\ref{sec:conv-anal}, we define the back transform
\begin{equation}
    F^\lambda(\rho) = (-E^\lambda)^*(-\rho) \label{eq:DFT-func-2}
\end{equation}
as a new functional that is convex and lower semicontinuous by construction.
It needs to be stressed here that although $F^\lambda = (\FDM^\lambda)^{**}$ by the two previous equations, this does \emph{not} necessarily mean that the two functionals agree, since $f^{**}=f$ \emph{only} holds if the functional is convex \emph{and} lower semicontinuous. It nevertheless generally holds that $F^\lambda(\rho)\leq \FDM^\lambda(\rho)$. But in any case we can get $E^\lambda(v)$ back from $F^\lambda(\rho)$ by
\begin{equation}
    E^\lambda(v)=-(F^\lambda)^*(-v). \label{eq:DFT-func-1}
\end{equation}
Now, if the convex $\FDM^\lambda(\rho)$ is also lower semicontinuous (Assumption~\ref{ass:DFT} below), this implies
\begin{equation}
    F^\lambda(\rho) = \FDM^\lambda(\rho).
\end{equation}
Unlike convexity, lower semicontinuity of $\FDM^\lambda(\rho)$ must be checked for every choice of topology.

As said before, in settings where $\FDM^\lambda(\rho)$ is not lower-semicontinuous,
the convex conjugate of $E^\lambda(v)$ does notably \emph{not} retrieve the original constrained-search functional $\FDM^\lambda(\rho)$ in Eq.~\eqref{eq:DFT-func-2}. Thus it looses connection to the variational principle in Eq.~\eqref{eq:E-def} and consequently also to the equivalent ground-state problem involving the Schrödinger equation.
This means that minimizers $\rho \in X$ in Eq.~\eqref{eq:E-var}
can occur where the convex conjugate $F^\lambda(\rho)$ of $E^\lambda(v)$ is finite, yet the constrained-search functional has $\FDM^\lambda(\rho)=+\infty$. If, on the other hand, $v$ in $E^\lambda(v)$ supports a ground state $\Psi$, then $\FDM^\lambda(\rho_\Psi)$ is also finite.
For this reason, it needs to be stressed that, albeit the requirement of lower semicontinuity for the ensemble constrained-search functional seems like a mathematical sophistry, it is a critical feature of the theory, and we will elaborate further on this on page~\pageref{page-lsc}.
The convex $F^\lambda(\rho)=\FDM^\lambda(\rho)$ on $X$ and the concave $E^\lambda(v)$ on $X^*$, as a conjugate pair, then serve as the backbone of a convex-analysis formulation of DFT.

Based on the above discussion, we 
formulate a separate assumption set for a consistent DFT framework. 

\begin{tcolorbox}[simple]
\begin{assumptions}\label{ass:DFT}
    $X$ and $X^*$ are such that the Hamiltonian $\hat H_v^\lambda$ is self-adjoint with a $v$-independent domain and bounded below, and the ensemble constrained-search functional $\FDM^\lambda(\rho)$ is lower semicontinuous. This implies $\FDM^\lambda(\rho)=F^\lambda(\rho)$.
\end{assumptions}
\end{tcolorbox}

To fulfill this assumption, we can think about Lieb's choice $L^1(\R^3)\cap L^3(\R^3)$ for the space of densities $X$, which indeed makes $\FDM^\lambda(\rho)$ lower semicontinuous~\cite{Lieb1983}. Yet, this space does not fulfill Assumption~\ref{ass:BS-reflexive}, which is not necessary for a consistent formulation of DFT but needed for our later constructions involving MY regularization. Instead, we suggest to switch to a more general yet reflexive space $X\supsetneq L^1(\R^3)\cap L^3(\R^3)$. This, on the other hand, can break lower semicontinuity of the ensemble constrained-search functional if we switch to a coarser topology. 

The dual pairing between potentials and densities in $X$ suggests a matching space for potentials that consists of the topological dual $X^*$. For $L^1(\R^3)\cap L^3(\R^3)$, the dual space is $L^{3/2}(\R^3)+L^\infty(\R^3)$. If instead $X\supsetneq L^1(\R^3)\cap L^3(\R^3)$, then $X^* \subsetneq L^{3/2}(\R^3)+L^\infty(\R^3)$. Interestingly, $L^{3/2}(\R^3)$ is exactly the boundary case in terms of $L^p$ spaces and the Hamiltonian fails to be self-adjoint on its usual domain for $p<3/2$ when the potentials can become too singular. This is the case for a potential $r^{-\alpha}$, $\alpha\geq 2$, where \emph{different} self-adjoint realizations on the punctured domain $\R^3\setminus\{0\}$ become possible. To ensure a self-adjoint and bounded-below Hamiltonian on the usual domain, we thus need to limit the potential space to $X^* \subseteq L^{3/2}(\R^3)+L^\infty(\R^3)$.

At this point, another problem appears due to this limitation. In certain cases, this space will not accommodate a given physical setting---for example, it might exclude singular Coulomb potentials and thus molecular systems in a clamped-nuclei or Born--Oppenheimer approximation. We can solve this problem by just including a fixed \emph{offset potential} from our previous class
\begin{equation}\label{eq:Kato-perturb}
    \vo \in L^{3/2}(\R^3)+L^\infty(\R^3).
\end{equation}
It can also be taken from any other potential class that still admits a self-adjoint and bounded-below Hamiltonian according to Assumption~\ref{ass:DFT}, including non-local potentials. In particular, $\vo$ can fulfill the task of a binding potential for $N$ electrons.
But one has to take careful note that such a potential can change the domain of the self-adjoint Hamiltonian if it is not from the class of Eq.~\eqref{eq:Kato-perturb}~\cite[Sec.~X.2]{reed-simon-2}. 
Instead of writing $E^\lambda(\vo+v)$, we redefine the energy functional as 
\begin{equation}
    E^\lambda(v) = \inf_{\Psi} \langle \Psi, \hat H^\lambda_{\vo+v} \Psi\rangle,
\end{equation}
including the fixed offset automatically. Note that the offset potential then must also enter the definition of the universal density functional so that it remains compatible to $E^\lambda(v)$,
\begin{equation}\label{eq:FDM-with-offset}
    F_\mathrm{DM}^\lambda(\rho) = \inf_{\hat\Gamma\mapsto\rho}\trace(\hat H^\lambda_{\vo}\hat \Gamma).
\end{equation}
While the inclusion of this offset potential does not change anything within the basic structure of DFT, which is the reason for not including it explicitly into the notation of the functionals, it cannot be ignored when solving the underlying equations, and we will highlight its occurrence at several places in the following sections. 

Whenever a minimizer can be found in the Hohenberg--Kohn variation principle of Eq.~\eqref{eq:E-var} with the functional $F_\mathrm{DM}^\lambda(\rho)$, we call this $\rho\in D$ a \emph{$v$-representable} density. This means, in other words, that there is a $v\in X^*$ such that the (ensemble) ground-state solution $\hat\Gamma$ to the Schrödinger equation with Hamiltonian $\hat H^\lambda_{\vo+v}$ yields $\hat\Gamma\mapsto\rho$ and ground-state energy $E^\lambda(v)=F_\mathrm{DM}^\lambda(\rho)+\langle v,\rho \rangle$.
We next write out the relation in Eq.~\eqref{eq:DFT-func-2} as the Lieb variational principle,
\begin{equation}
    \label{eq:LegendreFenchel}
    F^\lambda(\rho) = \sup_{v \in X^*} \{ E^\lambda(v) - \langle v,\rho \rangle \}.
\end{equation}
Since in general $F^\lambda(\rho) \leq \FDM^\lambda(\rho)$ and since a $v$-representable $\rho$ is a minimizer in Eq.~\eqref{eq:E-var} and must fulfill $\FDM^\lambda(\rho)=E^\lambda(v) - \langle v,\rho \rangle$, we see that $F^\lambda(\rho) = \FDM^\lambda(\rho)$ always holds for $v$-representable densities, even when $\FDM^\lambda(\rho)$ is not lower semicontinuous.
In what follows, we want to relate this property to the functional $F^\lambda(\rho)$ and its subdifferential.

The notions of sub- and superdifferentials serve an important purpose in the formulation of DFT.
If there exists an optimizer $v\in X^*$ for a given $\rho\in X$ in Eq.~\eqref{eq:LegendreFenchel}, 
then the following statements are all equivalent for the lower-semicontinuous $F^\lambda(\rho)$,
\begin{equation}
\begin{aligned}
v \in -\partial F^\lambda(\rho) &\iff \rho \in \partial E^\lambda(v) \\ & \iff  E^\lambda(v) = F^\lambda(\rho) + \langle v,\rho \rangle.
\label{eq:subdiff-equivalence}
\end{aligned}
\end{equation}
We can combine this into the inverse subdifferential relation,
\begin{equation}\label{eq:inv-subdiff-E-F}
    (-\partial F^\lambda)^{-1}(v) = \partial E^\lambda(v).
\end{equation}
Yet, it is by no means guaranteed that the set $-\partial F^\lambda(\rho)$ is non-empty for all $\rho\in D$.
But since $F^\lambda(\rho)$ is convex and lower semicontinuous by definition, it is known that its subdifferential is non-empty on a dense subset of $D\subset X$, the effective domain of the functional~\cite[Cor.~2.44]{Barbu-Precupanu}.

\label{page-lsc}
If $\FDM^\lambda(\rho)$ is lower semicontinuous, then $\FDM^\lambda(\rho) = F^\lambda(\rho)$, and what was said above also applies to $\FDM^\lambda(\rho)$.
In addition, in this case, each $\rho \in \partial E^\lambda(v)$
is an optimizer in the second line of Eq.~\eqref{eq:E-var} and therefore $\rho\in D$ . The convex set $\partial E^\lambda(v)\subseteq D$ then yields all possible ground-state densities for the potential $v\in X^*$.
This $\partial E^\lambda(v)$ will be empty if $v$ does not support a ground state for interaction strength $\lambda$, includes a single element if $v$ supports a nondegenerate ground state, and contains multiple elements in case of degeneracy. 
We may now identify the points at which $F^\lambda(\rho)$ has a non-empty subdifferential with the  ($v$-representable) ground-state densities of a Schrödinger equation. Hence, assuming a lower semicontinuous $\FDM^\lambda(\rho)$, it also holds that the set of $v$-representable densities is dense in the set of $N$-representable densities.
In this way, it can be highlighted in a different manner why the lower semicontinuity of the ensemble constrained-search functional is important. Else, with a $\FDM^\lambda(\rho)$ that is not lower semicontinuous, it is possible to find $v\in-\partial F^\lambda(\rho)$ when at the same time $\FDM^\lambda(\rho)=+\infty$, so this density has $\rho\in X\setminus D$ and would clearly be  not $v$-representable. This is why Assumption~\ref{ass:DFT} is important when dealing with the functionals $\FDM^\lambda(\rho)$ and $F^\lambda(\rho)$. See also the example given after the proposition below.

But it is also known since a long time that $v$-representability, even if it is secured for a dense set, is a problematic notion since most densities $\rho\in D$ will not have this property~\cite{ENGLISCH1983,vanLeeuwen2003-key-concepts}.
In many cases (including standard DFT), the set of $v$-representable densities is known only implicitly (exceptions are lattice systems~\cite{chayes1985density,Penz2021-GraphDFT,Penz2023-geom-deg} and particles on a one-dimensional torus~\cite{Sutter2024,CarvalhoCorso2025}).
Even worse, in the standard setting of DFT with density space $L^1 \cap L^3$, arbitrarily close to each $v$-representable density there lies a non-$v$-representable one, which makes $F^\lambda(\rho)$ discontinuous and non-differentiable~\cite{Lammert2007}.
Such an irregular behavior of the universal functional $F^\lambda(\rho)$ will create complicacies in the optimization procedures that DFT relies on. Further, full (interacting and non-interacting) $v$-representability is imperative for the construction of the KS iteration scheme (see Section~\ref{sec:ks}). If, on the other hand, $F^\lambda(\rho)$ would be differentiable, the derivative $v = -\nabla F^\lambda(\rho)$ would yield an immediate solution to the $v$-representability question, because this potential $v$ then exactly gives $\rho$ as a ground-state density by Eq.~\eqref{eq:subdiff-equivalence}.
We will see in Section~\ref{sec:MY-DFT} how MY regularization offers a way out of this dilemma.

The Hohenberg--Kohn theorem~\cite{Hohenberg1964} states that, for a $v$-representable density $\rho\in D$, the corresponding potential is unique up to the addition of a constant. In standard DFT, a rigorous proof of the Hohenberg--Kohn theorem~\cite{Garrigue2018HK} has been achieved for $v\in L^p(\R^3)+L^\infty(\R^3)$, with $p>2$, and it extends to closely related settings~\cite{garrigue2019hohenberg}. This leaves $3/2 \leq p \leq 2$ as undecided (the lower bound is given by the argument for self-adjoint Hamiltonians from before), although results on unique continuation~\cite{Jerison1985,Sogge1990} suggest that the Hohenberg--Kohn theorem holds also in this range. In non-standard DFT settings, like finite lattices or when currents are included, counterexamples are known~\cite{Penz2021-GraphDFT,Penz2023-Review-PartII}.

Combining the assumptions given in this section and in Section~\ref{sec:func-anal} (which will later serve as the basis for MY regularized DFT), we arrive at the following important result:
\begin{tcolorbox}[simple]
    \noindent
    \emph{Proposition}.
    Assumption~\ref{ass:BS-reflexive} together with Assumption~\ref{ass:DFT} with a continuously embedded potential space $X^* \subset L^{3/2}(\R^3)+L^\infty(\R^3)$ imply that every $v\in X^*$ supports a ground state---that is, for each $v\in X^*$, the superdifferential $\partial E^\lambda(v)$ is non-empty.
\end{tcolorbox}
This claim can be seen as follows. The choice $L^{3/2}(\R^3)+L^\infty(\R^3)$ as a potential space makes $E^\lambda(v)$ continuous~\cite[Th.~3.1(iii)]{Lieb1983}, implying that each continuously embedded $X^*\subset L^{3/2}(\R^3)+L^\infty(\R^3)$ gives a continuous energy functional $E^\lambda(v)$ as well. But a continuous and concave functional on $X^{*}$ admits a non-empty subdifferential belonging to $X^{\ast\ast}$ everywhere in the interior of its domain (here all $X^\ast$)~\cite[Prop.~2.36]{Barbu-Precupanu}. In general, $X^{**}\supseteq X$, but with Assumption~\ref{ass:BS-reflexive} we have reflexivity and thus $X=X^{**}$. By the additional requirement that $\FDM^\lambda(\rho)$ is lower semicontinuous, we have $F^\lambda(\rho)=\FDM^\lambda(\rho)$. Each $\rho\in\partial E^\lambda(v)$ is now in $D\subset X$ and through $v\in -\partial F^\lambda(\rho)=-\partial \FDM^\lambda(\rho)$ it is connected to an actual ground-state solution of the Schrödinger equation as argued above. This demonstrates the claim of the proposition.

\label{page-example}
We illustrate this subtle and important point with an example. Take the Hamiltonian from Eq.~\eqref{eq:Ham} with just a single particle and set $v_\mathrm{off}=v=0$. Then there is no bound state and thus $\partial E^\lambda(v)$ should be empty (note that the $\lambda$ parameter would not be needed for single particles since there are no interactions). Indeed, this system is incompatible with reflexivity and lower semicontinuity assumed together. If we give up on reflexivity and in contrast to Eq.~\eqref{eq:subdiff-dual} allow a superdifferential in $X^{**}$, then we can find an $x\in \partial E^\lambda(0)$ where $x\in X^{**} \setminus X$, which does not represent a ground state but instead corresponds to a scattering state.
Next, we keep reflexivity but give up on lower semicontinuity of $\FDM(\rho)$. We pick a minimizing sequence $\rho_i\to x$ in Eq.~\eqref{eq:E-var} consisting, for example, of normalized Gaussian-shaped densities $\rho_i \in D$ that get broader with increasing $i$, while $x \notin D$. Note that this means $x=0$, which cannot hold if $X$ includes the $L^1$ norm (and which would make it non-reflexive), since all $\rho_i$ were assumed normalized to the same number and clearly $N=\|\rho_i\|_{L^1}\to 0$ is a contradiction. But it is not necessarily a contradiction in a reflexive $X$ without the $L^1$ norm.
Now the chosen sequence fulfills $\lim\FDM(\rho_i)=0$ but at the same time $\FDM(x)=+\infty$, since $x\notin D$, so lower semicontinuity indeed cannot hold.

If, on the other hand, we add a fixed offset potential to the Hamiltonian that always guarantees existence of a ground state, for example, a harmonic oscillator potential $\vo(\rr)=|\rr|^2$, then the sequence $\rho_i$ of Gaussian-shaped densities from before cannot minimize $\FDM(\rho)$ any more. This is because the definition of $\FDM(\rho)$ includes the offset potential by Eq.~\eqref{eq:FDM-with-offset} and, as the densities get broader, the potential energy $\langle\vo,\rho_i\rangle$ diverges. Every potential that has $\lim_{|\rr|\to\infty}v_\mathrm{off}(\rr)=\infty$ guarantees existence of a ground state~\cite[XIII.47]{reed-simon-4} and we call it a \emph{confining potential}.

This section summarized the most important features of DFT in the language of functional and convex analysis. The reader may wonder if Assumption~\ref{ass:BS-reflexive} (or one of the  two stronger ones) and Assumption~\ref{ass:DFT} can  be satisfied simultaneously. After all, the classical setting of  DFT, with $X = L^1(\R^3)\cap L^3(\R^3)$ for densities and its dual $X^\ast = L^{3/2}(\R^3)+L^\infty(\R^3)$ for potentials, allows for Assumption~\ref{ass:DFT} but breaks reflexivity, a property not needed for a consistent formulation of DFT but necessary for our later results related to Moreau--Yosida regularization. 
One solution is to choose a bounded domain $\Omega$, which implies a continuous embedding $L^p(\Omega) \subseteq L^q(\Omega)$, $q<p$.
Given that the important result of lower semicontinuity due to \citet[Th.~4.4 and Cor.~4.5]{Lieb1983} holds with respect to the $L^1$ topology, it extends to each $L^p(\Omega)$, $p>1$. We can then choose any $X=L^p(\Omega)$ with $1<p<\infty$ with dual $X^* = L^{p^*}(\Omega)$, $1/p+1/p^{*}=1$, and thus have Assumptions~\ref{ass:BS-reflexive-unif-convex} and \ref{ass:DFT} simultaneously satisfied. By selecting $p=2$, we satisfy both Assumptions~\ref{ass:HS} and \ref{ass:DFT}. The previous works of \citet{eschrig2010t} and \citet{Kvaal2014} used this route, while many other works on the topic had to give up on at least one of the assumptions. 
Rather than switching to a bounded domain, we can use the offset potential $\vo$ to introduce a confining potential for the whole system and thus ensure that every $v\in X^*$ yields a ground-state density. This, at the same time, serves to break the full translational and rotational symmetry of the setting, which is necessary to allow for any localized ground state in the first place~\cite{Sutcliffe2012}.

\begin{tcolorbox}[simple]
    \noindent
    \emph{Summary}. To allow for any of Assumptions~\ref{ass:BS-reflexive}--\ref{ass:HS} together with \ref{ass:DFT} simultaneously, possible strategies are to switch to a bounded domain $\Omega \subset \R^3$ or to add a confining offset potential $\vo$.
\end{tcolorbox}

\subsection{Kohn--Sham method}
\label{sec:ks}

DFT reformulates the ground-state problem of quantum mechanics by replacing the many-particle wave function or density matrix as the central variable with the one-particle density, thereby significantly reducing the complexity of the description.
Yet, the usefulness of such a reduced description entirely depends on whether $F^\lambda(\rho)$ can be evaluated  accurately and efficiently.
The $F^0(\rho)$ of a non-interacting system amounts to just the kinetic energy and its conjugate functional $E^0(v)$ can be computed effectively by means of orbitals.
For the interacting case, however, $F^1(\rho)$ is in principle unavailable since its evaluation corresponds to a quantum version of an NP-hard problem~\cite{Verstraete2009}. 
The difference $F^1(\rho)-F^0(\rho)$, which relates to the exchange--correlation functional, has moreover turned out to be difficult to model in terms of $\rho$~\cite{Toulouse2022-chapter}.
Thus, although the exact DFT formulation is  very appealing by avoiding the full description by a wave function, further steps are needed to attain a useful reformulation. 
In the KS method~\cite{kohn1965self}, an auxiliary system of non-interacting electrons is introduced and its external potential $v_\rms$ is chosen in  such a way that it reproduces the same density as the interacting system. Yet, we will argue below that it is by no means clear that this approach is always possible.

The \emph{exchange--correlation potential} is defined as the potential that must be added to the external potential and the Hartree potential of the auxiliary, non-interacting system in order to get the same ground-state density as the interacting system. This means that the method requires $\rho\in D$ to be simultaneously interacting and non-interacting $v$-representable, which is equivalent to $\partial F^1(\rho)$ and $\partial F^0(\rho)$ both non-empty. We already noted on the general issues regarding $v$-representability in Section~\ref{sec:formalDFT}, and this additional requirement of simultaneous $v$-representability comes as a further restriction~\cite{Trushin2024}. Assuming for now that this is ensured, we can take $v\in-\partial F^1(\rho)$ and $v_\rms\in-\partial F^0(\rho)$ and define the exchange--correlation potential
\begin{equation}\label{eq:def-vxc}
    \vxc(\rho) = v_\rms - v - \vH(\rho).
\end{equation}
Here we already subtracted the Hartree potential
\begin{equation}
    \vH(\rho) = \nabla \EH(\rho) = \int \frac{\rho(\rr')}{|\cdot-\rr'|}\rmd\rr',
\end{equation}
which itself is the functional derivative of the Hartree mean-field energy
\begin{equation}
    \label{eq:Hartree-energy}
    \EH(\rho) = \frac 1 2 \iint \frac{\rho(\rr)\rho(\rr')}{|\rr-\rr'|}\rmd\rr\rmd\rr'.
\end{equation}
The total external potential that needs to be applied to the non-interacting system in order to get $\rho$ is then, not forgetting the previously introduced offset potential,
\begin{equation}\label{eq:vs}
    \vo + v_\rms = \vo + v + \vH(\rho) + \vxc(\rho),
\end{equation}
As was already noted, the assumption of (simultaneous interacting and non-interacting) $v$-representability comes as a severe limitation.
But putting this issue aside for the moment and assuming the existence of $\vxc(\rho)$ for all $\rho$ in question, Eq.~\eqref{eq:vs} can be translated into the KS self-consistent iteration scheme,
\begin{equation}\label{eq:KS1}
\arraycolsep=0pt
\begin{array}{cccccc}
    v_\rms & = v + \vH(&\rho&) + \vxc(&\rho&) \\
    \updownarrow && \updownarrow && \updownarrow & \\
    v_{i+1} & = v + \vH(&\rho_i&) + \vxc(&\rho_i&) 
\end{array}
\end{equation}
and
\begin{equation}\label{eq:KS2}
    \rho_{i+1}\in\partial E^0(v_{i+1}).
\end{equation}
The second step corresponds to solving the non-interacting Schrödinger equation with an external potential $v_{i+1}$ and determining a ground-state density $\rho_{i+1}$. If the Hamiltonian is defined on a bounded domain $\Omega\subset\R^3$ or contains a confining offset potential $v_\mathrm{off}$, this approach is always possible. A different convex-analytical argument for the non-emptiness of the superdifferential is given by our proposition building on Assumption~\ref{ass:BS-reflexive} together with Assumption~\ref{ass:DFT} and this was similarly used before~\cite[proof of Th.~12]{KSpaper2018}.
The two steps of Eqs.~\eqref{eq:KS1} and \eqref{eq:KS2} can be combined into a single step, either for the potentials or, as we will do here, for the densities,
\begin{equation}\label{eq:KS-1-step}
    \rho_{i+1}\in\partial E^0(v + \vH(\rho_i) + \vxc(\rho_i)).
\end{equation}
For initialization, we  choose an initial density $\rho_1$ or, equivalently, an initial potential $v_1$ for the auxiliary system that roughly leads to the desired ground-state density $\rho$.
Since the auxiliary system does not include any interactions between the particles, we choose (in the simplest case) its ground state as a Slater determinant and solve Eq.~\eqref{eq:KS2} by filling up the orbitals of a one-particle Schrödinger equation with external potential $\vo + v_{i+1}$.
If the procedure converges, we get
\begin{equation}
    \rho_i \to \rho, \qquad
    v_i \to v_\rms,
\end{equation}
so indeed both systems share the same ground-state density if, in the auxiliary system, an external potential $\vo+v_\rms$ is applied.
Finally, in order to conveniently retrieve the interacting energy $E^1(v)$ from a converged density and the expressions defined above, we can just use $E^0(v_\rms)=F^0(\rho)+\langle v_\rms,\rho\rangle$ and calculate
\begin{equation}\label{eq:KS-energy}\begin{aligned}
    E^1(v)&=F^1(\rho)+\langle v,\rho \rangle \\
    &=F^0(\rho) + F^1(\rho) - F^0(\rho) +\langle v-v_\rms+v_\rms,\rho \rangle \\
    &=E^0(v_\rms) + \Exc(\rho)+\EH(\rho) -\langle \vxc(\rho) +\vH(\rho),\rho \rangle \\
    &=E^0(v_\rms)+\Exc(\rho)-\EH(\rho)-\langle\vxc(\rho),\rho\rangle,
\end{aligned}\end{equation}
where we have introduced the \emph{exchange--correlation functional} $\Exc(\rho)=F^1(\rho)-F^0(\rho)-\EH(\rho)$. We remark that practical KS-DFT methods are usually based on approximating $\Exc(\rho)$ and assuming differentiability, we then have $\vxc = \nabla\Exc(\rho)$.

Note that this formulation of the KS method does not come with any guarantee of convergence. Indeed, in practice, an additional damping step~\cite{KARLSTROM_CPL67_348,cances2000can,cances2001}, like taking $\rho_i+\alpha(\rho_{i+1}-\rho_{i})$ with $\alpha<1$ as the next density, is introduced to facilitate convergence. A mathematical proof of convergence has only been achieved in the MY-regularized setting, which is in any case needed to avoid the $v$-representability problem, as will be discussed in Section~\ref{sec:epsKS-convergence}.

\section{Moreau--Yosida regularization in density-functional theory}
\label{sec:MY-DFT}

Having the stage set, we now present three different ways to introduce MY regularization into DFT, which are equivalent and thus have exactly the same outcome---namely, replacing the universal density functional $F^\lambda(\rho)$ with a differentiable $F_\eps^\lambda(x)$ for all $x\in X$. Which way is to be preferred, is up to the reader, but we think it is illustrative to put them all side to side. Some readers might find one or the other easier to understand or more or less conceptually appealing. In Section~\ref{sec:MY-total-energy}, we show that $F_\eps^\lambda(x)$ is the convex conjugate of an energy functional $E_\eps^\lambda(v)$ that includes the field energy of $v$ (if appropriately defined), while Section~\ref{sec:MY-mixing} identifies every $x\in X$ as a mix of internal and external densities.
In this section, we only require Assumption~\ref{ass:BS-reflexive}, but it is useful to also have Assumption~\ref{ass:DFT} with $F^\lambda=\FDM^\lambda$ fulfilled, so that no ambiguity about the respectively used universal functional arises.

\subsection{MY based on the universal density functional}
\label{sec:MY-univ}

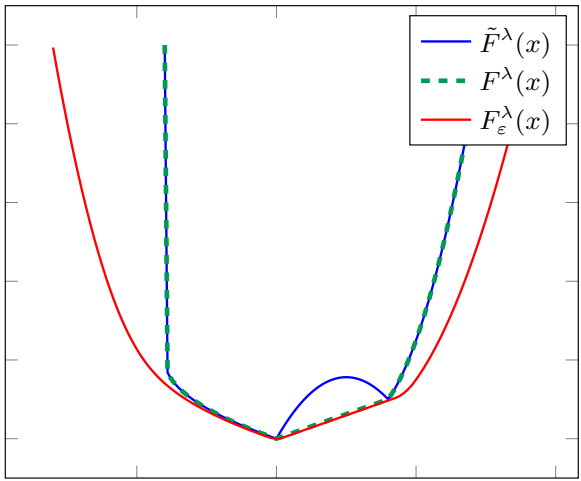
\begin{figure}[t]
\centering
\begin{tikzpicture}[scale=1.1]
    \draw[color=MidnightBlue,thick,->] (1.91,5) -- (1.91,5.4) node[right] {$+\infty$};
    \begin{axis}[restrict y to domain=0:10,xlabel={},ylabel={},xticklabels=\empty,yticklabels=\empty,xlabel={$x$},xlabel near ticks,xlabel shift=-5pt]
    \addplot[no markers,color=MidnightBlue,thick] table[x index=0,y index=1] {MY-example-data.dat};
    \addplot[no markers,color=ForestGreen,ultra thick,dashed] table[x index=0,y index=2] {MY-example-data.dat};
    \addplot[no markers,color=red,thick] table[x index=0,y index=3] {MY-example-data.dat};
    \legend{$\tilde F^\lambda(x)$,$F^\lambda(x)$,$F_\eps^\lambda(x)$}
    \end{axis}
    \draw[MidnightBlue,fill=MidnightBlue] (1.91,0) circle (1.5pt);
    \draw[color=MidnightBlue,very thick] (1.91,0) -- (6,0) node[midway,above,color=MidnightBlue] {$D$};
    \draw[color=MidnightBlue,very thick,dotted] (6,0) -- (6.5,0);
\end{tikzpicture}
\caption{Example of a pure-state functional $\tilde F^\lambda(x)$, the convex $F^\lambda(x)$, and its MY regularization $F_\eps^\lambda(x)$. The unregularized functionals jump to $+\infty$ for all $x$ outside their effective domain $D$, but the regularized functional remains finite everywhere.}
\label{fig:MY-example}
\end{figure}

The most straightforward way to introduce MY regularization into DFT follows from its definition in Eq.~\eqref{eq:MY}: By  directly applying the method to the non-differentiable (but convex and lower semicontinuous) $F^\lambda(\rho)$, yielding a regularized universal density functional $F_\eps^\lambda(x)$ that is well-defined and finite for all $x\in X$, 
\begin{equation}\label{eq:Feps}
    F_\eps^\lambda(x) = \inf_{\rho\in X}\left\{ F^\lambda(\rho) + \frac{1}{2\eps}\|x-\rho\|_{X}^2\right\},
\end{equation}
where $\eps > 0$ is the smoothing parameter.
An illustration of a regularized functional in one dimension together with the potentially non-convex $\tilde F^\lambda(x)$ and convex $F^\lambda(x)$ is shown in Figure~\ref{fig:MY-example}.
The regularized functional  has a well-defined functional derivative (see \ref{item:diff-f_eps} from the properties of MY regularization in Section~\ref{sec:MY}) and the regularization is lossless in the sense that it conserves the overall infimum of the functional (see \ref{item:inf}).

We next introduce the corresponding energy functional $E_\eps^\lambda(v)$ as the convex conjugate of $F_\eps^\lambda(x)$,
\begin{equation}\label{eq:Eeps-def}
    E_\eps^\lambda(v) = -(F_\eps^\lambda)^*(-v) = \inf_{x\in X}\{F_\eps^\lambda(x) + \langle v,x \rangle\}.
\end{equation}
Note that the resulting $E_\eps^\lambda(v)$ is \emph{not} the MY regularization of $E^\lambda(v)$. Instead, the following relation holds (from Eq.~\eqref{eq:feps-conj}, where the different sign is due to Eq.~\eqref{eq:DFT-func-1}),
\begin{equation}
    E_\eps^\lambda(v) = E^\lambda(v) -\frac{\eps}{2}\|v\|^2_{X^*}.
\end{equation}
Since we subtract the norm-square of $v$ from the concave $E^\lambda(v)$, the MY energy $E_\eps^\lambda(v)$ is \emph{strongly concave}, which implies strict concavity. Conversely, it holds that the Legendre--Fenchel transform of a strictly concave energy functional is always differentiable.

With $v = -\nabla F_\eps^\lambda(x)$ we now have a \emph{unique} representing potential $v\in X^*$ for every $x\in X$. We here write $\nabla F_\eps^\lambda$ instead of  $\partial F_\eps^\lambda$ because we know that the regularized functional is differentiable. This then automatically yields a well-defined and unique mapping from `densities' (elements of $X$) to potentials (elements of $X^*$) without any $v$-representability issue. We are not even limited to the $N$-representability set $D\subset X$ any more, and Section~\ref{sec:MY-mixing} will help to make sense of the generalized densities $x\in X$. Uniqueness of the potential also means we automatically have a Hohenberg--Kohn-type result. 

In the Hilbert-space setting (Assumption~\ref{ass:HS}) we further obtain a \emph{quantitative} bound that shows how the difference between two potentials is bound by the difference between the corresponding densities. This comes in form of the Lipschitz estimate from property~\ref{item:frechet}:
\begin{equation}\label{eq:nablaF-Lipschitz}
    \|\nabla F_\eps^\lambda(x)-\nabla F^\lambda_\eps(y)\|_{X^*} \leq \eps^{-1} \|x-y\|_X .
\end{equation}
This bound is lost in the limit $\eps\to 0$, in agreement with the fact that, in \emph{standard} DFT, two densities that lie very close to each other can come from vastly different potentials~\cite{garrigue2021-pot-to-gs}, making density--potential inversion ill-posed and thus notoriously difficult.
A similar bound is also possible in more general Banach spaces if the MY regularization is adapted to the `modulus of convexity' of the space~\cite{Penz2025-pMY}.
Furthermore, MY regularization allows for a well-defined formulation of the KS iteration procedure (Section~\ref{sec:regKS}).

\subsection{MY based on the total energy functional}
\label{sec:MY-total-energy}

A different viewpoint starts from the concave ground-state energy functional $E^\lambda(v)$ and introduces a \emph{strictly concave} energy $E_\eps^\lambda(v)$ by just subtracting $\eps\phi(v)=\frac{\eps}{2}\|v\|^2_{X^*}$:
\begin{equation}\label{eq:Eeps-strictly-convex}
    E_\eps^\lambda(v) = E^\lambda(v)-\frac{\eps}{2}\|v\|^2_{X^*}.
\end{equation}
Convex conjugation of this strictly concave functional then  yields the regularized density functional of Eq.~\eqref{eq:Feps} as shown by Eq.~\eqref{eq:feps-conj},
\begin{equation}
    F_\eps^\lambda(x) = (-E_\eps)^*(-x) = \sup_{v\in X^*}\left\{ E_\eps^\lambda(v)-\langle v,x\rangle \right\}.
\end{equation}
We have thus defined MY regularization starting from the energy functional instead of the universal density functional. But talking about the total energy allows to give this manipulation a physical interpretation. For a given external potential $v\in X^*$, we introduce a \emph{field energy functional}
\begin{equation}\label{eq:Efield}
    \Efield(v)=-\frac{\eps}{2}\|v\|^2_{X^*},
\end{equation}
where the minus sign that makes the functional concave, just like the concavity of the matter energy functional $E^\lambda(v)$. Of course, to retrieve the correct physical electrostatic \emph{field energy}, $\frac{\eps_0}{2}\|\nabla v\|^2_{L^2}$, an appropriate choice of Banach space $X^*$ must be made; see Section~\ref{sec:Hartree-autoreg}. We then introduce the new energy functional (that could be interpreted as the total energy of matter \emph{and} field)
\begin{equation}
    E_\eps^\lambda(v) = E^\lambda(v)+\Efield(v),
\end{equation}
which corresponds to Eq.~\eqref{eq:Eeps-strictly-convex}.
In this way, MY regularization arises \emph{naturally} by attaching more physical relevance to the external potential and by assigning to it an energy that has to be taken into account in the total energy functional.

Regarding the negative sign in Eq.~\eqref{eq:Efield}, remember that in Eq.~\eqref{eq:Eeps-def} it is $F_\eps^\lambda(x) + \langle v,x \rangle$ that is varied in order to find the ground state. There, the additional term is $\tfrac 1 {2\eps}\|x-\rho\|^2_X$ as seen in Eq.~\eqref{eq:Feps} and using the relation motivated in Eq.~\eqref{eq:mixing} below this is equal to $\tfrac \eps 2 \|v\|^2_{X^*}$, precisely the expression for the  (positive) field energy.
Viewed in this manner, it is the energy content of the external field that regularizes $F^\lambda(\rho)$. Conceptually, this is an extremely attractive viewpoint, since it explains the regularization as a physical effect.

\subsection{MY based on density--potential mixing}
\label{sec:MY-mixing}

Remember that we defined the $N$-representability set $D\subset X$ in Eq.~\eqref{eq:D-def} as the set of all $\rho \in X$ where $\FDM^\lambda(\rho)<\infty$, known as the effective domain of $\FDM^\lambda(\rho)$. For the sake of argument, we now suppose Assumption~\ref{ass:DFT} under which $\FDM^\lambda(\rho)=F^\lambda(\rho)$.
Now even for $\rho\in D$, we can have issues with $\partial F^\lambda(\rho)$. First, $\partial F^\lambda(\rho)$ may be empty. In such cases,  no potential $v \in X^\ast$ exists such that $v\in -\partial F^\lambda(\rho) \iff\rho \in \partial E^\lambda(v)$, implying that $\rho$ is not $v$-representable.  
Also, at $v$-representable densities a small variation of $\rho$ can lead outside the set of $v$-representable densities~\cite{Lammert2007,garrigue2021-pot-to-gs}. There are further examples on small lattice systems where even large variations of $v$ lead to almost (or even exactly) the same ground-state density $\rho$ (and thus give counterexamples to the Hohenberg--Kohn theorem)~\cite{Penz2021-GraphDFT,Penz2024-SpinLatticeDFTPerspective}. 

It would be better altogether if we could ignore this difficult notion of ``$v$-representability'' and instead are able to give a physical interpretation to all elements $x\in X$. This is achieved with the inverse duality map $J^{-1}\colon X^*\to X$ as the canonical way of mapping potentials to densities. For any $v$-representable $\rho\in\partial E^\lambda(v)$ coming from $v\in X^*$ define the \emph{mixed density}
\begin{equation}\label{eq:mixing}
    x = \rho -\eps J^{-1}(v).
\end{equation}
This definition is unique only if a Hohenberg--Kohn-type result is available so that the corresponding $v\in -\partial F^\lambda(\rho)$ is unique, but even without uniqueness we can make this assignment and will simply allow multiple such combinations.
Choosing $\eps > 0$ small means that in order to get the generalized density, we only mix in a small quantity of the corresponding potential $v$. We note that the resulting $x$ does not need to be positive, normalizable, or have a finite kinetic energy, only the $v$-representable density $\rho$ has all those properties.
This is immediately visible from the definition in Eq.~\eqref{eq:mixing}, where $x$ is seen to be the sum of the electronic density $\rho$ and $-\eps J^{-1}(v)$.
With $v\in X^*$ as an \emph{external} potential with assigned density $-\eps J^{-1}(v)$ that gets added to the \emph{internal} electronic density of a quantum system, one could call $x$ a `bidensity'. It has been given different names before, such as `depot' (for density--potential), `pseudodensity'~\cite{Kvaal2014}, or `quasidensity'~\cite{Penz2023-Review-PartI}. We will not commit to a singular nomenclature here, and instead refer to it as a general element $x\in X$ or a `mixed density' when its decomposition like in Eq.~\eqref{eq:mixing} matters.

One physically motivated possibility is to choose the spaces such that $J$ corresponds to solving the Poisson equation of electrostatics and $\eps J^{-1}(v)=-\eps\Delta v$. Then $\rho_\mathrm{ext}=\eps J^{-1}(v)$ can be physically interpreted as just the external density that enters the Poisson equation to yield the electrostatic potential $v$. The mixed density is then the total observable density of the system, $x=\rho-\rho_{\mathrm{ext}}$ (the minus sign is needed since we have used the convention that the electronic density is positive, $\rho\geq 0$). Interestingly, even the MY parameter $\eps$ then receives a physical interpretation as the vacuum permittivity. This choice was previously suggested in the context of density--potential inversion~\cite{Penz2023-MY-ZMP}; it will be covered in more detail in Section~\ref{sec:Hartree-autoreg} and has already found multiple use in ameliorating the Kohn--Sham method~\cite{Görling1999,Heaton-Burgess2007,Callow2020-improving,Callow2020-screening}, albeit not with the connection to the topology of density and potential space.

Since $\rho \in \partial E^\lambda(v) = \partial (F^\lambda)^*(-v)$ from Eq.~\eqref{eq:DFT-func-1} (not forgetting a minus from the chain rule) we note that remarkably, Eq.~\eqref{eq:mixing}  reproduces Eq.~\eqref{eq:diff-feps-conj} from the properties of the MY regularization. We just have to use $E_\eps^\lambda(v)=-(F_\eps^\lambda)^*(-v)$ from Eq.~\eqref{eq:Eeps-def} and the homogeneity of the duality map to get
\begin{equation}\label{eq:x-dE-Jinv}
    x \in \partial E_\eps^\lambda(v) = \partial E^\lambda(v) - \eps J^{-1}(v),
\end{equation}
which is just the superdifferential of Eq.~\eqref{eq:Eeps-strictly-convex}.
Since $\partial E_\eps^\lambda$ has a well-defined inverse, $-\nabla F_\eps^\lambda$, it follows that for any $x\in X$ we can find a $v\in X^*$ such that the representation in Eq.~\eqref{eq:mixing} holds. In other words, \emph{every} $x\in X$ can be represented as a mixed density.
The regularizing effect of the mixing for $F^\lambda(\rho)$ is now easily explained: If $v\neq v'$ lead to (almost) the same ground-state densities $\rho\approx\rho'$, then this will not be true any more for $x = \rho -\eps J^{-1}(v)$ and $x' = \rho' -\eps J^{-1}(v')$ due to the additional external term.

Also the proximal map receives a simple interpretation. If we set $v=-\nabla F_\eps^\lambda(x)$ in Eq.~\eqref{eq:mixing} and solve for $\rho$, it follows from property \ref{item:diff-f_eps} in Section~\ref{sec:MY} that
\begin{equation}\label{eq:dens-prox}
    \rho = x - \eps J^{-1}(\nabla F_\eps^\lambda(x)) = \prox_{\eps F^\lambda}(x).
\end{equation}
Hence the internal density is just the proximal point of the mixed density. The consequence for DFT is that, to be able to work with the MY-regularized functionals, we also have to switch from internal densities $\rho\in D$ to general, mixed densities $x\in X$. The usual internal densities can then always be recovered as the proximal point or by performing the limit $\eps \to 0$, where the limit procedure is well-defined through the properties of the MY regularization.

\section{Density--potential inversion}
\label{sec:inversion}

The limit procedure in property \ref{item:diff-f_eps-convergence} of the MY regularization in Section~\ref{sec:MY} that says $\lim_{\eps\to 0}\nabla F_\eps^\lambda(x) \in \partial F^\lambda(x)$ will be the basis for a method to calculate the corresponding external potential from a given ground-state density. The method itself is already known since a long time as the ``ZMP method''~\cite{Zhao1994}, but our derivation as well as the accompanying interpretation is only feasible by resorting to MY regularization. We also add a more speculative subsection that makes an interesting and previously unexplored connection to the Dyson equation.
In this section we will require Assumption~\ref{ass:BS-reflexive} but sometimes, when it is mentioned, demand higher regularity of the involved spaces for additional results. Assumption~\ref{ass:DFT} ($F^\lambda=\FDM^\lambda$) is further necessary in order to stay connected to solutions of the Schrödinger equation as explained in Section~\ref{sec:formalDFT}.

As a motivation, we start with the variational problem for the universal density functional from Eq.~\eqref{eq:LegendreFenchel},
\begin{equation}
    F^\lambda(\rho) = \sup_{v\in X^*} \{E^\lambda(v) - \langle v, \rho\rangle\},
\end{equation}
whose stationarity condition $\partial E^\lambda(v) + \rho \ni 0$ gives $v\in -\partial F^\lambda(\rho)$ as an optimizer if $\rho$ is assumed $v$-representable. This approach to density--potential inversion was first applied to atoms by Colonna and Savin~\cite{Colonna1999} in 1999 and by Teale, Coriani and Helgaker to molecules~\cite{Teale2009,Teale2010a,Teale2010b} in 2009. It is sometimes called ``Lieb optimization'', ``Lieb maximization''~\cite{Teale2010a,Teale2010b,Lieb-Book-Trygve} or the ``dual problem''~\cite{Garrigue2022} in the literature.
The method of \citet{Wu2003} turns this principle into a practical constrained optimization problem for the KS system.
Convergence of Lieb optimization can be facilitated by making the functional strictly concave by subtracting $ \frac{\varepsilon}{2}\Vert v \Vert^2_{X^*}$, which already corresponds to MY regularization,
\begin{equation}\begin{aligned}
    F_\eps^\lambda(\rho) &= \sup_{v\in X^*} \{E_\eps^\lambda(v)- \langle v, \rho\rangle \}\\
    &= \sup_{v\in X^*} \left\{E^\lambda(v) - \frac{\varepsilon}{2}\Vert v \Vert^2_{X^*} - \langle v, \rho\rangle\right\}.
\end{aligned}\end{equation}
The stationarity condition for an optimizer $v_\eps$ is then
\begin{equation}
    0 \in \partial E^\lambda(v_\varepsilon) - \varepsilon J^{-1}(v_\varepsilon) - \rho.
\end{equation}
We use Eq.~\eqref{eq:x-dE-Jinv} to translate this to $\rho \in \partial E_\eps^\lambda(v_\varepsilon)$ which is equivalent to $v_\eps=-\nabla F_\eps^\lambda(\rho)$ by using the analogous relation to Eq.~\eqref{eq:inv-subdiff-E-F} for the regularized functionals. By property \ref{item:diff-f_eps} from Section~\ref{sec:MY}, we can connect $v_\eps$ to the proximal map,
\begin{equation}\label{eq:veps-prox-intro}
    v_\eps = \eps^{-1}J(\prox_{\eps F^\lambda}(\rho)-\rho).
\end{equation}
This equation then serves as the basis for MY density--potential inversion as an iterative procedure involving the proximal point of $\rho$, to find $v_\eps$, which in the limit $\eps\to 0$ gives the representing potential of $\rho$. The method will be more properly derived from MY regularization in the following section.

\subsection{Proximal-point iteration and relation to ZMP}
\label{sec:prox-ZMP}

The inverse problem of finding the corresponding potential $v$ for a given ground-state density $\rho$ is central to DFT and  to achieve this, we `only' have to evaluate the subdifferential of $F^\lambda(\rho)$ (with a minus sign, see Eq.~\eqref{eq:subdiff-equivalence}),
\begin{equation}\label{eq:v-dF}
    v\in-\partial F^\lambda(\rho).
\end{equation}
As mentioned before, it is not guaranteed that this subdifferential is non-empty even for $\rho\in D \subset X$ and, in non-standard settings for DFT (like DFT on finite lattices~\cite{Penz2021-GraphDFT}), it may even be multi-valued. For the regularized functional, no such problems occur, and we can evaluate
\begin{equation}\label{eq:v-dFeps}
    v_\eps=-\nabla F_\eps^\lambda(\rho)
\end{equation}
for any $\rho\in D$ and even for $\rho\in X \setminus D$. Of course, this does in general not yield the same potential as Eq.~\eqref{eq:v-dF}, but it does if $\rho=\prox_{\eps F^\lambda}(\rho)$ by property \ref{item:diff-f_eps} in Section~\ref{sec:MY} or as $\eps\to 0$ by property \ref{item:diff-f_eps-convergence}.
We therefore know how the functional derivative in Eq.~\eqref{eq:v-dFeps} relates back to the non-regularized form, but we may wonder how it can be computed.

Noting relation \ref{item:diff-f_eps} for the proximal map and introducing $p=\prox_{\eps F^\lambda}(\rho)$, we obtain
\begin{equation}
    \nabla F_\eps^\lambda(\rho) = \eps^{-1}J(\rho-p) \in \partial F(p).
\end{equation}
which by Eq.~\eqref{eq:subdiff-equivalence} gives
\begin{equation}
    p \in \partial E^\lambda(\eps^{-1}J(p-\rho)).
\end{equation}
Now we have the unknown $p\in D$ on both sides, so this form lends itself to an iteration scheme. Beginning with a guess $p_1$, we iterate
\begin{equation}
    \label{eq:prox-iter}
    p_{i+1} \in \partial E^\lambda(\eps^{-1}J(p_i-\rho)).
\end{equation}
Due to the possibility of degeneracy in the ground state, the next step $p_{i+1}$ is possible non-unique, which creates a potential issue that has not been explored up to date.
If converged, this process yields the proximal point $p=\prox_{\eps F^\lambda}(\rho)$ and we retrieve Eq.~\eqref{eq:veps-prox-intro},
\begin{equation}\label{eq:veps-prox}
    v_\eps = -\nabla F_\eps^\lambda(\rho) = \eps^{-1}J(p-\rho), 
\end{equation}
where the resulting potential $v_\eps$ depends on the choice of the smoothing parameter $\eps$.
Repeating this iterative scheme for a sequence of decreasing $\eps>0$, we can extrapolate to the limit $\eps=0$ and obtain $v_\eps \rightharpoonup v$. Under the minimal  Assumption~\ref{ass:BS-reflexive}, convergence  $v_\eps \rightharpoonup v$ is weak; under the stronger Assumption~\ref{ass:BS-reflexive-unif-convex}, convergence is strong. Also by virtue of \ref{item:diff-f_eps-convergence},  the resulting potential $v$ is the element of $\partial F^\lambda(\rho)$ with the smallest norm. 

The benefit of this method is that we do not need to evaluate the (usually unknown) $F^\lambda(\rho)$, relying instead on $\partial E^\lambda(v)$ by solving the Schrödinger equation for the ground state. 
This density-potential inversion procedure using MY regularization and its relation to the closely related ZMP method~\cite{Zhao1994} was previously detailed in \citet{Penz2023-MY-ZMP}.

\subsection{Moreau--Yosida in inverse Kohn--Sham}

While the procedure from the preceding section remains a hard task for interacting particles, density--potential inversion is often considered for the non-interacting KS system ($\lambda=0$) and then typically specified as the ``inverse Kohn--Sham method''. Starting with a given ground-state density $\rho$ for some external potential $v$ in an interacting system that is also non-interacting $v$-representable, the Kohn--Sham potential must satisfy
\begin{equation}
    v_\rms \in -\partial F^0(\rho).
\end{equation}
We can approach this potential according to Eq.~\eqref{eq:veps-prox} with
\begin{equation}
v_\rms = \lim_{\varepsilon\to 0} \frac{1}{\varepsilon} J\left(\prox_{\eps F^0}(\rho)-\rho\right),
\end{equation}
where the proximal map can be evaluated by using the iteration in Eq.~\eqref{eq:prox-iter}. To obtain $\rho\in\partial E^0(v)$, we solve the non-interacting Schrödinger equation, introducing orbitals.

A more intuitive viewpoint for this reasoning is to compare the corresponding minimization of the MY regularization and the very KS idea. Since $\rho$ is by assumption a minimizer of $F^0(\cdot) + \langle v_\rms, \cdot\rangle$, and $\prox_{\eps F^0}(\rho)$ is the minimizer of $F^0(\cdot) + \tfrac{1}{\eps}\Vert \cdot -\rho\Vert^2$, we must have simultaneously
\begin{alignat}{10}
    &0\in \partial F^0(\rho) \,\,&+\,\, &v_\rms, \\
    &0\in \partial F_\eps^0(\prox_{\eps F^0}(\rho)) \,\,&+\,\, &\frac{1}{\eps} J (\prox_{\eps F^0}(\rho) -\rho ).
\end{alignat}
Motivated by the fact that $\prox_{\eps F^0}(\rho)\to \rho$ as $\eps\to 0$, one can then under technical assumptions prove that indeed $v_\rms$ corresponds to the limit of $\tfrac{1}{\eps} J (\prox_{\eps F^0}(\rho) -\rho )$ as the equations above suggest.

Now exactly this method was proposed and implemented as a density--potential inversion scheme by \citet{Zhao1994} (and was later termed ``ZMP method''), where the choice of spaces corresponds to homogeneous Sobolev spaces for potentials and their duals for densities (also see Section~\ref{sec:Hartree-autoreg} and an upcoming publication for periodic domains like in Section~\ref{sec:MY-periodic}).
But in the original publication the method was developed \emph{ad hoc} and only proven to work empirically. Only quite recently it was shown~\cite{Penz2023-MY-ZMP} that the proximal-point iteration with $\eps\to 0$ relates closely to the ZMP method for the usual Sobolev spaces and that it also reproduces a method by \citet{vanleeuwen1994exchange} if employed on a Hilbert space with $J=J^{-1}=\id$. It is further equivalent to yet another density--potential inversion method by \citet{Kumar2020}, where different choices for their penalty function $S[\rho]$ would correspond to different topologies for $X$. The method has been analyzed numerically~\cite{Herbst2025,Bohle2026} (see also Section~\ref{sec:MY-periodic} here) and adds itself to a long list of KS inversion techniques~\cite{Shi2021}. More than that, it shows that MY regularization offers a mathematical framework that allows for an elegant unification of numerous density--potential inversion methods.

\subsection{Resolvent form and Dyson-like equation}

Instead of establishing a practical scheme for density--potential inversion, this section will be devoted to an interesting reformulation of the basic relation $v=-\nabla F_\eps^\lambda(x)$ that becomes possible for a regularized functional.
Starting from Eq.~\eqref{eq:diff-feps-conj} for the universal density functional $F^\lambda$ and using Eq.~\eqref{eq:inv-subdiff} we get
\begin{equation}
    (\nabla F_\eps^\lambda)^{-1}(v) = \partial (F^\lambda)^*(v) + \eps J^{-1}(v).
\end{equation}
Even though $\nabla F_\eps^\lambda(x)$ is single-valued, its inverse is in general a set-valued map, so the equation above is actually between sets.
From Eq.~\eqref{eq:DFT-func-1} we get (not forgetting the minus from the chain rule) $\partial (F^\lambda)^*(-v)=\partial E^\lambda(v)$ and we apply this to the function above after switching $v\mapsto-v$, where we also use that $J^{-1}$ is homogeneous. Then for $v=-\nabla F_\eps^\lambda(x)$ (or, more directly, from Eq.~\eqref{eq:x-dE-Jinv}) we have
\begin{equation}\label{eq:dF-inv-x}
    (\nabla F_\eps^\lambda)^{-1}(-v) = \partial E^\lambda(v) - \eps J^{-1}(v) = x.
\end{equation}
Finally, we invert the whole expression, arriving at a resolvent-type formula that can be applied to all elements from $X$, the density space, to yield the corresponding potential,
\begin{equation}
    -\nabla F^\lambda_\eps = (\partial E^\lambda - \eps J^{-1})^{-1}.
\end{equation}
Note that in the Hilbert-space setting we already saw a resolvent form for the proximal map in property \ref{item:prox-resolvent} in Section~\ref{sec:MY}.
By transforming Eq.~\eqref{eq:dF-inv-x} into an iterative procedure to solve for $v$, we arrive at
\begin{equation}\label{eq:Dyson}
    -\nabla F_\eps^\lambda = -\eps^{-1}J + \eps^{-1}J\circ\partial E^\lambda \circ (-\nabla F_\eps^{\lambda}).
\end{equation}
Here we used `$\circ$' to denote the concatenation of operations and, unsurprisingly, this result is equivalent to the proximal-point iteration in Eq.~\eqref{eq:prox-iter} if we combine it with Eq.~\eqref{eq:veps-prox}. But writing it in the form above exposes a striking parallel to the Dyson equation~\cite{fetter-walecka2012}, where $-\nabla F_\eps^\lambda$ takes the role of the interacting Green's function (dressed propagator), $-\eps^{-1}J$ that of the free Green's function (bare propagator), and $-\partial E^\lambda$ (with an additional minus to get the sign right) would be the self-energy. Applying $-\eps J^{-1}$ from the left and $(-\nabla F_\eps^\lambda)^{-1} = \partial E_\eps^\lambda$ from the right yields
\begin{equation}
    \partial E_\eps^\lambda = \partial E^\lambda - \eps J^{-1},
\end{equation}
which is just Eq.~\eqref{eq:x-dE-Jinv} again and shows another path to arrive at Eq.~\eqref{eq:Dyson}. In our setting of MY applied to DFT, the mapping $-\eps^{-1}J$, that gets defined by the topology of the space $X$ itself, gives the `bare' connection between densities and potential. Compared to this, $-\nabla F_\eps^\lambda$ is a `dressed' way how to connect the quantities, where the dressing occurs through the presence of particles the potential can interact with. Lastly, $-\partial E^\lambda$ derives from the functional $E^\lambda(v)$, the energy of the matter system. This last element introduces a possible ambiguity into the formulation that needs to be considered if degeneracy occurs and $\partial E^\lambda(v)$ is a multi-valued set. Nevertheless, this reformulation allows for an interesting alternative viewpoint that links it to Green's function methods and diagrammatic expansions.
We hope that this can serve as a starting point for future investigations, either for concrete approximations to the potential $-\nabla F_\eps^\lambda$ or into how to effectively link this back to the corresponding unregularized form $-\partial F^\lambda$.

\section{Regularized Kohn--Sham theory}
\label{sec:regKS}

We already noted in Section~\ref{sec:ks} that the requirement of having all involved densities simultaneously interacting and non-interacting $v$-representable puts a severe limitation on the possibility of a rigorous formulation of the Kohn--Sham method. This limitation can also influence numerical applications of the method, since it means that basis-set limits are ill-defined. The MY regularized functional $F^\lambda_\eps(x)$ offers a way how to avoid representability issues since the functional is differentiable and every $x\in X$ becomes representable through $-\nabla F^\lambda_\eps(x)$. The corresponding regularized Kohn--Sham method was first formulated in \citet{Kvaal2014} for a Hilbert-space setting and later extended to Banach spaces~\cite{KSpaper2018}. In this section, we not only give its basic formulation, but also show how usual exchange--correlation can be translated to the new method. Finally, we repeat the main steps of a convergence proof for the method with an additional damping step.
As usual, we require Assumption~\ref{ass:BS-reflexive} but have to move to the Hilbert-space setting (Assumption~\ref{ass:HS}) and even finite-dimensional spaces for the convergence proof. Since in the KS method we ought to stay connected to the Schrödinger equation and related KS equation, also Assumption~\ref{ass:DFT} ($F^\lambda=\FDM^\lambda$) is necessary.

\subsection{Iteration scheme}
\label{sec:epsKS}

We will build directly on the iteration scheme introduced in Section~\ref{sec:ks} but, with the help of MY regularization, can completely avoid the critical issues within the usual formulation of the KS method. As a reminder, we were able to introduce a well-defined exchange--correlation potential $\vxc(\rho)$ only if $\rho$ is simultaneously interacting and non-interacting $v$-representable, i.e., at every density $\rho$ that occurs within the KS iteration both subdifferentials $\partial F^1(\rho)$ and $\partial F^0(\rho)$ must be non-empty. From what we learned about MY regularization, this can be directly guaranteed if instead we switch to the regularized functionals $F^1_\eps(x)$ and $F^0_\eps(x)$. However, we then need to make sense of the resulting regularized KS iteration procedure. So let us define in exact analogy to Eq.~\eqref{eq:def-vxc} for any $y \in X$ with $v(y)=-\nabla F^1_\eps(y)$ and $v_\rms(y)=-\nabla F^0_\eps(y)$,
\begin{equation}\label{eq:def-vxceps}\begin{aligned}
    \vxceps(y) &= v_\rms(y) - v(y) - \vH(y) \\
    &= \nabla F^1_\eps(y) - \nabla F^0_\eps(y) - \vH(y) \\
    &= \nabla(\underbrace{F^1_\eps(y) - F^0_\eps(y) - \EH(y)}_{\textstyle \Exceps(y)}).
\end{aligned}\end{equation}
We now write $y$ instead of $\rho$ to highlight that this is an arbitrary element in $X$ that does not need to have the properties of an actual density (see also Section~\ref{sec:MY-mixing}).
We can now define the same, self-consistent iteration procedure as in Eqs.~\eqref{eq:KS1} and \eqref{eq:KS2}, where we can use Eq.~\eqref{eq:x-dE-Jinv} (which is equivalent to Eq.~\eqref{eq:diff-feps-conj}) for the superdifferential of the $E^0_\eps$ functional. Let the external potential that acts on the fully interacting system now be fixed as $\vo+v$ with $v=-\nabla F^1_\eps(x)$ (as before, the offset $\vo$ will be included into the functionals), then iterate starting from an initial guess,
\begin{align}
    & v_{i+1} = v + \vH(x_i) + \vxceps(x_i), \label{eq:KS1-eps} \\
    & x_{i+1}\in\partial E^0_\eps(v_{i+1}) = \partial E^0(v_{i+1}) - \eps J^{-1}(v_{i+1}). \label{eq:KS2-eps}
\end{align}
In solving for the next $x_{i+1}$ in Eq.~\eqref{eq:KS2-eps}, we rely on exactly the same non-interacting Schrödinger equation that allows an orbital construction as in the usual KS procedure, we just have to shift the resulting density by $- \eps J^{-1}(v_{i+1})$ as the density--potential mixing dictates.
If the procedure converges (see Section~\ref{sec:epsKS-convergence} for a convergence proof that applies to finite-dimensional $X$ if an additional damping step is added), we get
\begin{equation}
    x_i \to x, \qquad
    v_i \to v_\rms,
\end{equation}
and the resulting $x$ can then be mapped to the ground-state density with the proximal map according to Eq.~\eqref{eq:dens-prox}. But now an important difference to the usual KS procedure appears that was already apparent in the formulation of the regularized KS iteration: The method aims at a shared $x\in X$ for the full and auxiliary system, and not at the internal ground-state density. This means that when $\vo+v_\rms$ is applied to the auxiliary system we get the same $x=\rho_\rms-\eps J^{-1}(v_\rms)$ as when $\vo+v$ is applied to the full system. Therefore, in general $\rho\neq\rho_\rms$ and instead we have the ground-state densities
\begin{align}
    &\rho = x+\eps J^{-1}(v) = \prox_{\eps F^1}(x), \label{eq:rho-prox}\\
    &\rho_\rms = x+\eps J^{-1}(v_\rms) = \prox_{\eps F^0}(x). \label{eq:rhos-prox}
\end{align}
That it must hold $\rho,\rho_\rms\in D$ serves as a consistency check for the method or can be used as an \emph{a posteriori} correction.
To get an expression for the ground-state energy, we can rely on the analogous expression to Eq.~\eqref{eq:KS-energy},
\begin{equation}
    E^1_\eps(v)=E^0_\eps(v_\mathrm{s})+\Exceps(x)-\EH(x)-\langle\vxceps(x),x\rangle,
\end{equation}
that we just need to shift as in Eq.~\eqref{eq:feps-conj},
\begin{equation}
    E^1(v)=E^1_\eps(v)+\frac{\eps}{2}\|v\|^2_{X^*}.
\end{equation}
Note, however, that while we recover the usual ground-state energy $E^1(v)$ this way, it does not mean that the regularized Kohn--Sham method will yield the same orbitals as the usual method. This is clear since the orbitals in regularized Kohn--Sham will produce a density $\rho_\rms$ that is different to the interacting density $\rho$, and instead the generalized densities $x=\rho-\eps J^{-1}(v)=\rho_\rms-\eps J^{-1}(v_\rms)$ agree.

\subsection{Transforming exchange--correlation potentials}
\label{sec:transform-xc}

While this reformulation of the KS iteration scheme with the help of MY regularization is conceptually already very appealing, the question remains how it relates back to the unregularized setting.
Having again $v=-\nabla F^1_\eps(x)$, in the first KS step that is Eq.~\eqref{eq:KS1-eps} we rely on an exchange--correlation potential for the regularized scheme at every step $x_i$ for which so far only the defining relation Eq.~\eqref{eq:def-vxceps} was given.
More concretely, we want to know how $\vxceps$ relates to the usual $\vxc$ or any approximation to it. For this we take Eqs.~\eqref{eq:rho-prox} and \eqref{eq:rhos-prox} and combine them into
\begin{equation}\label{eq:vxceps-from-rhos}
    \vxceps(x) = \eps^{-1}J(\rho_\rms-x) - \eps^{-1}J(\rho-x) - \vH(x).
\end{equation}
Note that if $X$ is a Hilbert space then $J$ is a linear mapping and the expression above can be simplified. Within the regularized KS iteration we already have access to $\rho_\rms$ in every step as an element of $\partial E^0(v)$, so we just need the corresponding $\rho\in\partial E^1(v)$ in order to evaluate the above equation. Since we of course want to avoid solving the full problem, the $\vxc$ from standard KS-DFT (or any approximation to it) enters the picture, since it allows getting the same density from the auxiliary system by adding just the usual exchange--correlation potential, $\rho\in\partial E^0(v+\vH(\rho)+\vxc(\rho))$. Together with $v = \eps^{-1}J(\rho-x)$, we can derive a self-consistent iteration that works similar to KS itself,
\begin{equation}
\arraycolsep=0pt
\begin{array}{cccccccc}
    \rho & \in \partial E^0(\eps^{-1}J(&\rho&-x)+\vH(&\rho&)+\vxc(&\rho&)) \\
    \updownarrow && \updownarrow && \updownarrow && \updownarrow & \\
    \rho_{i+1} & \in \partial E^0(\eps^{-1}J(&\rho_i&-x)+\vH(&\rho_i&)+\vxc(&\rho_i&)).
\end{array}
\end{equation}
If this iteration is integrated into the regularized KS scheme from Section~\ref{sec:epsKS} together with determining $\vxceps(x)$ from Eq.~\eqref{eq:vxceps-from-rhos}, we can recycle any given exchange--correlation approximation for the regularized setting.
It must be added that this transformation of exchange--correlation potentials for the regularized KS method was not implemented to date.

\subsection{Convergence proof}
\label{sec:epsKS-convergence}

Now that a rigorous formulation of the KS method has been achieved, the question of convergence of the method to the desired ground-state density, when using the exact exchange--correlation functional, remains to be answered. First results on convergence of the usual KS method (Section~\ref{sec:ks}) required a $v$-representability assumption and only showed convergence in energy~\cite{Wagner2013,WagnerFollowUp}. With the help of MY regularization the problem of $v$-representability could be avoided and by the additional inclusion of a damping step a full convergence proof was achieved at least for a setting where $X$ is a (finite-dimensional) Euclidean space~\cite{KS_PRL_2019,PRLerrata,kvaal2022-chapter}. This setting means that Assumption~\ref{ass:HS} holds, and we slightly generalize the result to any finite-dimensional Hilbert space $X$ in the following presentation.

\begin{figure}[ht]
\centering
\begin{tikzpicture}[scale=.8]
    \draw[->] (-.5,0) -- (7,0)
        node[right] {$x$};
    \draw[->]
        (0,-0.2) -- (0,6.5);
    
    \fill (0,0) circle [radius=2pt] node[below=5pt] {$x_i$}; 
    \fill (1,0) circle [radius=2pt] node[below=5pt] {$x_{i+1}$};
    \fill (6,0) circle [radius=2pt] node[below=2pt] {$x'_{i+1}$};
    \draw[dashed] (1,-.2) -- (1,6);
    
    \draw (-.5,5.125) parabola bend (1,4) (3,6) node[below right] {$\tfrac{1}{2\eps} \|x-x_{i+1}\|^2_X + m_i$};
    
    \draw (-.5,5) -- (3.5,1) node[right] {$F^1_\eps+v$};
    
    \draw[dotted] (-.5,5.5) -- (3.5,1.5) node[right] {$F^1+v$};
    
    \draw[dashed] (-.5,4.5) -- (7,4.5) node[right] {$e_i$};
    \draw[dashed] (-.5,4) -- (7,4) node[right] {$m_i$};
    \draw[dashed] (-.5,3.5) -- (7,3.5) node[right] {$e_{i+1}$};
\end{tikzpicture}
\caption{Illustration of one iteration step in the regularized KS scheme.}
\label{fig:epsKS}
\end{figure}
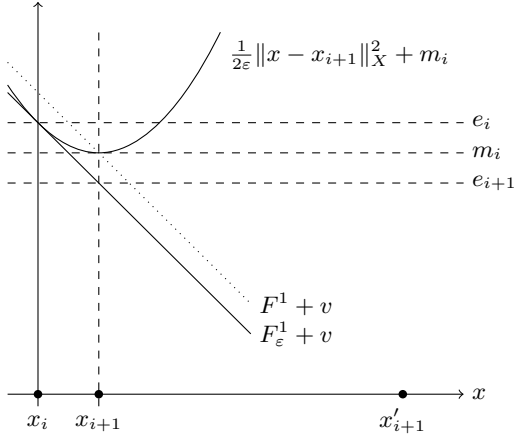

Implementations of the KS method regularly come with additional steps like damping that facilitate convergence. Such a damping step must also be included here, which means that in Eq.~\eqref{eq:KS2-eps} not the full step from $x_i$ to $x_{i+1}$, but one with a reduced step length is applied. This step length will be such that one reaches exactly the apex of the regularization parabola that is tangentially aligned to $F^1_\eps+v$ at $x_i$, as illustrated in Figure~\ref{fig:epsKS}. Yet, it must be stressed that such a choice of step-length does not yield a practical guideline, since it requires knowledge about the fully interacting functional. Its sole purpose here is to be able to establish a mathematical convergence proof. An actual implementation of the KS method must decide about the step length from the available information from the previous steps. Different schemes with an \emph{a priori} or adaptive choice of step length can be constructed as well~\cite{cances2001,KSpaper2018,CDFT-paper} and their convergence rates have been compared~\cite{CDFT-paper,KS_PRL_2019}. We also note on a strong parallel to the proximal-point algorithm~\cite{Parikh2014-prox}, where the minimum of $f(x)$ is found by iterating $x_{i+1}=\prox_{\eps f}(x_i)$. By the definition of the proximal map, Eq.~\eqref{eq:def-prox}, this corresponds to jumping to the apex of the full regularization function $\eps\phi(x-x_i)$ attached to $x_i$ in every step. The important difference to the regularized Kohn--Sham method is that the method fixes the step direction in the previous step (by solving the non-interacting system with the exchange--correlation potential) and then jumps to the apex of the regularization parabola restricted to this direction.

The regularized KS scheme with additional damping step is then given by the iteration
\begin{align}
    & v_{i+1} = v + \vH(x_i) + \vxceps(x_i), \label{eq:KS1-eps-damping} \\
    & x_{i+1}'\in\partial E^0_\eps(v_{i+1}) = \partial E^0(v_{i+1}) - \eps J^{-1}(v_{i+1}) \label{eq:KS2-eps-damping} \\
    & \quad\text{and set}\quad y_i=(x_{i+1}'-x_i)/\|x_{i+1}'-x_i\|_X, \nonumber\\
    & x_{i+1} = x_i + \tau_i y_i, \label{eq:KS3-eps-damping}\\
    & \quad\text{where}\quad \tau_i=-\eps\langle\nabla F^1_\eps(x_i)+v,y_i\rangle. \nonumber
\end{align}
In the following, we provide most steps and explanations from the original proof~\cite{KS_PRL_2019} that was subsequently revised due to an error in the first version~\cite{PRLerrata}. For additional details we refer to the original publications.

From the first step and using the definition of $\vxceps$ in Eq.~\eqref{eq:def-vxceps} we have that
\begin{equation}\label{eq:epsKS-F1-F0}
    \nabla F^1_\eps(x_i)+v = v_{i+1}+\nabla F^0_\eps(x_i).
\end{equation}
The iteration is converged if either side of this equation is zero, which by the inverse subdifferential relation, Eq.~\eqref{eq:inv-subdiff}, means that $x_i \in \partial E^1_\eps(v)$ and also $x_i \in \partial E^0_\eps(v_{i+1})$ (the interacting and the non-interacting system share the same $x$ as in Eqs.~\eqref{eq:rho-prox} and \eqref{eq:rhos-prox}). In order to track progression of the method, we define
\begin{equation}\label{eq:epsKS-ei}
    e_i = F_\eps^1(x_i) + \langle v,x_i \rangle.
\end{equation}
In case of correct convergence $x_i\to x\in \partial E^1_\eps(v)$ we then have $\lim e_i = E^1_\eps(v)$, which equals the ground-state energy of the interacting system minus $\frac{\eps}{2}\|v\|^2_{X^*}$. We take the expression of Eq.~\eqref{eq:epsKS-ei} that is differentiable and convex in $x_i$ and rewrite its directional derivative in the chosen direction $x_{i+1}'-x_i$ with the help of Eq.~\eqref{eq:epsKS-F1-F0},
\begin{equation}
    \langle \nabla F^1_\eps(x_i)+v, x_{i+1}'-x_i \rangle = \langle v_{i+1}+\nabla F^0_\eps(x_i), x_{i+1}'-x_i \rangle.
\end{equation}
Notice that $x_{i+1}' \in \partial E^0_\eps(v_{i+1})$ and $x_i\in \partial E^0_\eps(-\nabla F^0_\eps(x_i))$ by definition, so from Section~\ref{sec:MY} we can use the strong monotonicity property \ref{item:strong-mon} of $\partial E^0_\eps(v) = \partial (F^0_\eps)^*(-v)$ (note the change of sign) to get
\begin{equation}\label{eq:epsKS-estimate}\begin{aligned}
    \langle \nabla F^1_\eps(x_i)+v, x_{i+1}'-x_i \rangle &\leq -\eps\|v_{i+1}+\nabla F^0_\eps(x_i)\|^2_{X^*} \\
    &= -\eps\|\nabla F^1_\eps(x_i)+v\|^2_{X^*}.
\end{aligned}\end{equation}
This means that by a step in the direction $x_{i+1}'-x_i$, which is parallel to $y_i$, we can always decrease the energy, if the step length is not taken too large. For this reason we limit the step length to $\tau_i$ such that we just reach the apex of the regularization parabola (at value $m_i$). We get this value from equating the directional derivatives of $F^1_\eps(x_i)+v$ and the regularization parabola at $x_i$, which yields
\begin{equation}\label{eq:epsKS-step-equal}
    \langle \nabla F^1_\eps(x_i)+v, y_i \rangle = -\frac{1}{\eps}\|x_{i+1}-x_i\|_X = -\frac{\tau_i}{\eps}
\end{equation}
and guarantees $e_{i+1}\leq m_i < e_i$. Since the energy is bounded below and thus strictly decreasing, we already showed convergence in energy.
By determining $e_i-m_i$ from the regularization parabola and using Eq.~\eqref{eq:epsKS-step-equal} we get
\begin{equation}
    \frac{\tau_i^2}{2\eps} = \frac{1}{2\eps}\|x_{i+1}-x_i\|_X^2 = e_i-m_i\leq e_i-e_{i+1}\to 0
\end{equation}
that shows $\tau_i\to 0$.
Taking Eqs.~\eqref{eq:epsKS-estimate} and \eqref{eq:epsKS-step-equal} together gives the estimate
\begin{equation}\label{eq:epsKS-conv-to-v}
    \frac{\tau_i}{\eps} \|x_{i+1}'-x_i\|_X \geq \eps\|\nabla F^1_\eps(x_i)+v\|^2_{X^*}.
\end{equation}
If we can be sure that $\|x_{i+1}'-x_i\|_X$ is bounded this means that $-\nabla F^1_\eps(x_i) \to v$.

In order to show convergence in densities and potentials, a compactness argument must be used. The revised proof~\cite{PRLerrata} uses the fact that $F^1_\eps(x)$ increases like $\|x\|^2_X$ asymptotically outside of its effective domain $D$ as a result of MY regularization. For $X$ finite-dimensional, which critically enters here, all norms are equivalent and thus $D$ is bounded since it only includes normalized densities. Since $e_i=F_\eps^1(x_i) + \langle v,x_i \rangle < e_1$ from the result that the $e_i$ are strictly decreasing, we arrive at $(x_i)_i$ bounded. Again, the finite dimensionality of $X$ is critical in the next step, because it guarantees that a bounded sequence always has convergent subsequences. This means $(x_i)_i$ has (possibly infinitely many) convergent subsequences of which we exemplary take $x_{\alpha(i)}\to z$. Since we are in a Hilbert-space setting (Assumption~\ref{ass:HS}), property \ref{item:frechet} holds and $\nabla F^\lambda_\eps$ is Lipschitz continuous.
Now the first step Eq.~\eqref{eq:KS1-eps-damping} asserts
\begin{equation}\begin{aligned}
    v_{\alpha(i)+1} &= v + \vH(x_{\alpha(i)}) + \vxceps(x_{\alpha(i)}) \\
    &= v + \nabla F^1_\eps(x_{\alpha(i)}) - \nabla F^0_\eps(x_{\alpha(i)}),
\end{aligned}\end{equation}
so by continuity $v_{\alpha(i)+1} \to v + \nabla F^1_\eps(z) - \nabla F^0_\eps(z)$ and the subsequence is bounded. By the same argument as before also the sequence given by $x_{\alpha(i)+1}'\in\partial E^0_\eps(v_{\alpha(i)+1})$ is bounded which means $\|x_{\alpha(i)+1}'-x_\alpha(i)\|_X$ is bounded and so we really have $\lim_{i\to\infty}(-\nabla F^1_\eps(x_{\alpha(i)})) = -\nabla F^1_\eps(z) = v$ from Eq.~\eqref{eq:epsKS-conv-to-v}. This already implies $\lim_{i\to\infty} v_{\alpha(i)+1} =  - \nabla F^0_\eps(z)$, so we have found a potential for the non-interacting KS system reproducing $z\in\partial E^1_\eps(v)$. Taking a different subsequence $x_{\beta(i)}$ can then yield another element from $\partial E^1_\eps(v)$ as well as its corresponding KS potential.

\section{Auto-regularization in mean-field theories}
\label{sec:autoreg}

In certain, specialized settings, like the Hartree-approx\-i\-mation to DFT or DFT with induced magnetic fields (Maxwell--Schrödinger DFT), the special form of the additional energy terms leads to automatic MY regularization. Interestingly, in the first case this is not on the side of the universal density functional, as in Eq.~\eqref{eq:Feps}, but affects the ground-state energy $E^\lambda(v)$. In the case of Maxwell--Schrödinger DFT we can apply Legendre--Fenchel transformations to two different arguments and thus have the possibility to define four different functionals from convex conjugation, in some analogy to the potentials of thermodynamics that are also linked by Legendre--Fenchel transformations. There, it turns out that two of those functionals can even be linked to a certain form of double MY regularization called Lasry--Lions regularization.

\subsection{Hartree auto-regularization}
\label{sec:Hartree-autoreg}

Through a special choice for the potential space previously mentioned in Sections~\ref{sec:MY-total-energy} and \ref{sec:MY-mixing}, we can achieve automatic regularization of the energy functional in the Hartree approximation of DFT.
This choice is the homogeneous Sobolev space~\cite{mazya-book,galdi-book,ortner2012},
\begin{equation}\label{eq:hom-Sobolev}
	X^* = \{u\in L^1_\mathrm{loc}(\R^3) \mid \nabla u \in L^2(\R^3:\R^3) \},
\end{equation}
that must be taken modulo constants in order to make $\|u\|_{X^*} = \sqrt{\eps_0}\|\nabla u\|_{L^2}$ a valid norm. This includes the customary ``up to an additive constant'' already into the definition of the potential space. Note that we also built the vacuum permittivity $\eps_0$ ($=1/(4\pi)$ in atomic units) into the definition of the norm for reasons that will soon become evident. This space is actually a Hilbert space with inner product $\langle u,v \rangle_{X^*} = \eps_0\langle \nabla u,\nabla v \rangle_{L^2}$, so we have Assumption~\ref{ass:HS} fulfilled. Applying partial integration in the $L^2$ inner product gives
\begin{equation}\label{eq:Hartree-auto-partial}
    \langle u,v \rangle_{X^*} = -\eps_0 \int u(\rr)\Delta v(\rr)\rmd\rr = \langle u, J^{-1}(v) \rangle,
\end{equation}
where the last term stems from the Riesz representation Eq.~\eqref{eq:Riesz-rep} applied to $X^*$ and means that $u\in X^*$ is applied to $J^{-1}(v)\in X$. But this allows us to write the inverse duality map as
\begin{equation}\label{eq:J-Poisson}
    J^{-1}(v) = -\eps_0\Delta v,
\end{equation}
so evaluating $v=J(x)$ is equivalent to solving the Poisson equation $-\eps_0\Delta v = x \in X$. Strictly speaking, the integral in Eq.~\eqref{eq:Hartree-auto-partial} makes only sense if $\Delta v$ is still an $L^2$ function, but we can still take Eq.~\eqref{eq:J-Poisson} in the sense of distributions. This implies that a general element $x\in X$ can be a distribution. 
Since $J$ is the canonical mapping between densities and potentials, and is itself fully defined by the topology (norm) of the space, we needed to include $\eps_0$ in the norm's definition already to get the units right (although in atomic units one would not notice).
Now for regular distributions $x\in X$ (that can be represented by an $f\in L^1_\mathrm{loc}(\R^3)$ as $\langle u,x \rangle = \int u(\rr)f(\rr)\rmd\rr$) we know the solution of $-\eps_0\Delta v = x$ to be exactly the Hartree potential (in this context also called the Riesz potential; note again that in atomic units $4\pi\eps_0=1$),
\begin{equation}\label{eq:Riesz}
    v(\rr) = \int \frac{f(\rr')}{|\rr-\rr'|}\rmd\rr' = \vH(f)(\rr).
\end{equation}
Thus we have found an explicit expression for the duality map, $J(x)=\vH(f)$, when $x\in X$ is a regular distribution represented by $f\in L^1_\mathrm{loc}(\R^3)$. Note that we usually use the symbols for distributional densities `$x$' and their representing functions `$f$' for elements of $X$ interchangeably, so we could have also written $J(x)=\vH(x)$.
Further note that since the potentials are defined modulo a constant it must also hold $\langle u,x \rangle=\langle u+c,x \rangle$ which implies $\int f(\rr)\rmd\rr=0$, i.e., all integrable $f$ representing an $x\in X$ must integrate to zero. This may seem problematic if we think of densities $f$ as normalized to a certain particle number, but it can be compensated by the introduction of a fixed reference density and shifting $X$ to an affine space with the same dual $X^*$.
Now take the norm in $X^*$ with one-half in front,
\begin{equation}
    \frac{1}{2}\|v\|_{X^*}^2 = \frac{\eps_0}{2}\|\nabla v\|_{L^2}^2 = \frac{\eps_0}{2} \int |\nabla v(\rr)|^2 \rmd\rr,
\end{equation}
and we can recognize this as the electrostatic field energy.
We can equivalently rewrite the norm in $X$ with the dual norm as in Eq.~\eqref{eq:J-def},
\begin{equation}\label{eq:dens-dens}
\begin{aligned}
    &\frac{1}{2}\|v\|_{X^*}^2 = \frac{1}{2}\|x\|_{X}^2 = \frac{1}{2}\langle J(x),x \rangle \\
    &= \frac{1}{2}\iint \frac{f(\rr)f(\rr')}{|\rr-\rr'|}\rmd\rr\rmd\rr' = \EH(f),
\end{aligned}
\end{equation}
which retrieves the Hartree mean-field energy from Eq.~\eqref{eq:Hartree-energy} that is also the energy of the field induced by the charge distribution $f\in X$. Through these identifications, it becomes apparent that this choice of density and potential space is a very natural one for DFT and we believe that it can have a future importance for the formulation of DFT.
Indeed, the use of similar spaces in DFT was already suggested around 2012, based on an appearance in regularization of optimized-effective-potential methods~\cite{Heaton-Burgess2007}, but never published.
Still, several important mathematical questions are left to be studied in this setting, especially the self-adjointness and boundedness below of the Hamiltonian and lower semicontinuity of the ensemble constrained-search functional (Assumption~\ref{ass:DFT}). Like the summary of Section~\ref{sec:formalDFT} states, to make the the last point even possible to hold together with Assumption~\ref{ass:BS-reflexive}, one could switch to a bounded domain or include a confining offset potential $\vo$.

A very basic approximation for an interacting system is the Hartree functional
\begin{equation}\label{eq:F-Hartree-autoreg}
    \FHartree(\rho) = F^0(\rho) + \EH(\rho) = F^0(\rho) + \frac{1}{2}\|\rho\|_{X}^2,
\end{equation}
which becomes strongly (and thus strictly) convex through the addition of the Hartree mean-field energy.
We are now prepared to show the auto-regularization of the corresponding energy functional from the Legendre--Fenchel transform (the attentive reader might already have noticed that a strictly convex functional $\FHartree(\rho)$ implies a MY-regularized convex conjugate functional). With $v=J(x)$ we get
\begin{equation}\label{eq:EHartree}
\begin{aligned}
    &\EHartree(v) = -(\FHartree)^*(-v) \\[0.3em]
    &= \inf_\rho \left\{\FHartree(\rho) + \langle v,\rho \rangle \right\} \\
    &= \inf_\rho \left\{ F^0(\rho) + \frac{1}{2}\|\rho\|_{X}^2 + \langle v, \rho \rangle \right\} \\
    &= \inf_\rho \left\{ F^0(\rho) + \frac{1}{2}\|\rho\|_{X}^2 + \langle v, \rho \rangle + \frac{1}{2}\|v\|_{X^*}^2 \right\} - \frac{1}{2}\|v\|_{X^*}^2 \\
    &= \inf_\rho \left\{ F^0(\rho) + \frac{1}{2}\|\rho+x\|_{X}^2 \right\} - \frac{1}{2}\|v\|_{X^*}^2 \\
    &= F^0_{\eps=1}(-x) - \frac{1}{2}\|v\|_{X^*}^2 = F^0_{\eps=1}(-J^{-1}(v)) - \frac{1}{2}\|v\|_{X^*}^2.
\end{aligned}
\end{equation}
We can note several things here. In the third line we recognize that the considered energy content of the system is the (non-interacting) kinetic energy $F^0(\rho)$ of the system, the potential energy from coupling to the external $v$, but also the energy of the induced field $\EH(\rho)$. This sums up to give the self-interaction error of the Hartree approximation. The regularization parameter can be read off directly from the fifth line of Eq.~\eqref{eq:EHartree} as $\eps=1$. Without including the vacuum permittivity into the norm we can have $\eps=\eps_0$, but then the duality map does not directly map densities to potentials any more. The new argument of the regularized functional is $-x = -J^{-1}(v) = \eps_0\Delta v$ and can of course be negative as a general element in $X$. Note that now the energy functional $\EHartree(v)$ is the result of MY regularization, not the universal density functional as before. Even though we subtract $\frac{1}{2}\|v\|_{X^*}^2$, which reminds of the way MY regularization was introduced in Section~\ref{sec:MY-total-energy}, the energy functional does not necessarily become strictly concave (yet it must be concave by construction) since $F^0_\eps(-x)$ is convex. Since $\rho = \nabla \EHartree(v)$ is now unique, there arises no ambiguity in the ground-state density.

\subsection{Maxwell--Schrödinger DFT}
\label{sec:mdft}

Maxwell--Schrödinger DFT~\cite{tellgren2018-MDFT} is an extension of standard DFT with electromagnetic fields and internal currents that also considers induced magnetic fields. In this respect it is similar to the Hartree method discussed before in Section~\ref{sec:Hartree-autoreg}, adding a current-current integral instead of the density-density integral of Eq.~\eqref{eq:dens-dens}. Since the currents couple to an external vector potential, we include those into the description, just like in current DFT~\cite{Penz2023-Review-PartII}.
The vector potentials live in a special homogeneous and divergence-free (Coulomb gauged) Sobolev space that is similar to the previously mentioned homogeneous Sobolev space,
\begin{equation}
    Y^* = \{ \aa\in L^2(\R^3,\R^3) \mid \nabla\times\aa \in L^2(\R^3,\R^3), \nabla\cdot\aa = 0 \}.
\end{equation}
This space must further be taken modulo constant vector fields in order to make $\|\aa\|_{Y^*} = \mu_0^{-1/2}\|\nabla\times\aa\|_{L^2}$ a valid norm. Here, we built the vacuum permeability $\mu_0$ ($=4\pi/c^2$ in atomic units) into the norm of the space for vector potentials and it is again a Hilbert space. Since $\bb=\nabla\times\aa$ describes a magnetic field, $\frac{1}{2}\|\aa\|^2_{Y^*} = \frac{1}{2\mu_0}\|\bb\|^2_{L^2}$ exactly corresponds to the magnetostatic field energy. The corresponding space for current densities is then its dual space $Y$, where the duality map $J:Y\to Y^*$ then means solving the Maxwell equation
\begin{equation}
    -\Delta\aa=\nabla\times\bb=\mu_0\jj.
\end{equation}
It was possible to rewrite this as a set of component-wise Poisson equations since in the application of the vector Laplacian $\Delta\aa = \nabla (\nabla \cdot \aa) - \nabla\times(\nabla\times\aa)$ the divergence term vanishes for $\aa\in Y^*$ due to the chosen Coulomb gauge.
The current space $Y$ includes distributional vector fields and the norm for regular distributions $\jj\in Y$ features a current-current integral,
\begin{equation}
    \frac{1}{2}\|\aa\|_{Y^*}^2 = \frac{1}{2}\|\jj\|_Y^2 = \frac{\mu_0}{2}\iint \frac{\jj(\rr)\cdot\jj(\rr')}{4\pi|\rr-\rr'|}\rmd\rr\rmd\rr'.
\end{equation}

After setting the functional-analytic stage, let us introduce the coupled Maxwell--Schrödinger system. We redefine the kinetic-energy operator so that it includes vector potentials $\aa\in Y^*$ through minimal coupling,
\begin{equation}
    \hat T_{\aa} =-\frac 1 2 \sum_{j=1}^N (-\rmi\nabla_j+\aa)^2.
\end{equation}
In contrast to the original publication~\cite{tellgren2018-MDFT}, we do not include spin degrees-of-freedom, since this keeps the notation more concise but does not change the main argument about auto-regularization. We now define a Hamiltonian $\hat H^{\lambda=1}_{v,\aa}$ like in Eq.~\eqref{eq:Ham} just replacing $\hat T$ with $\hat T_{\aa}$. We further always choose the fully interacting case with $\lambda=1$ here and will thus suppress this parameter in the notation. The ground-state energy for a potential pair $(v,\aa)\in X^*\times Y^*$ is then
\begin{equation}
    E(v,\aa) = \inf_\Psi \langle \Psi ,\hat H_{v,\aa} \Psi \rangle,
\end{equation}
which can equivalently be defined with variation over density matrices. This shows that this functional is again concave in $v$, but has no special convexity property in $\aa$, a fact related to diamagnetism~\cite{tellgren2018-MDFT}.
We define the corresponding constrained-search functional over ensembles~\cite{Vignale1987,Vignale1988,Laestadius2014},
\begin{equation}
    F(\rho,\jp) = \inf_{\hat\Gamma\mapsto(\rho,\jp)} \trace(\hat H_{0,0}\hat\Gamma),
\end{equation}
that is now defined on $X\times Y$, the space of pairs of densities and paramagnetic currents. The notation $\hat\Gamma\mapsto(\rho,\jp)$ means that in the variation all density matrices that map to a given one-particle density $\rho$ and paramagnetic current $\jp$ are considered.
We then have
\begin{equation}
    E(v,\aa) = \inf_{\substack{\rho\in X\\ \jp\in Y}}\left\{F(\rho,\jp) + \langle v+\tfrac{1}{2}\aa^2,\rho \rangle + \langle\aa,\jp\rangle \right\}.
\end{equation}
One immediately sees that the density now additionally couples to $\tfrac{1}{2}\aa^2$, which means that the resulting energy functional is also not concave in $\aa$. This can easily be repaired by substituting $u=v+\tfrac{1}{2}\aa^2$, which makes the energy functional jointly concave in $u$ and $\aa$. But note that this construction requires $\aa^2\in X^*$, the space of potentials, which only holds when the spaces $X$ and $Y$ have a special compatibility property~\cite{CDFT-paper}.

Without requiring joint concavity, the energy functional for Maxwell--Schrödinger DFT is defined as
\begin{equation}\label{eq:EM-def}
    \EM(v,\AA) = \inf_{\aalpha\in Y^*} \left\{ E(v,\aalpha+\AA) + \frac{1}{2}\|\aalpha\|_{Y^*}^2 \right\}.
\end{equation}
Here and in the following, we exactly follow the notation of \citet{tellgren2018-MDFT} in order to allow for direct comparison. The stationarity condition for this energy functional then gives the Schrödinger equation with Hamiltonian $\hat H_{v,\aalpha+\AA}$ coupled to the Poisson equations for all components of the induced field $\aalpha\in Y^*$,
\begin{equation}
    -\Delta\aalpha = \mu_0 J^{-1}(\aalpha) = \mu_0\jj_\Psi.
\end{equation}
Here, $\jj_\Psi$ is the physical current from the solution of the Schrödinger equation. Next, we define $\bEM$ by subtracting the field energy $\frac{1}{2}\|\AA\|^2_{Y^*}$. Also substituting $\aa=\aalpha+\AA$, we see that we have arrived at a formula in close analogy to the Hartree auto-regularization in Eq.~\eqref{eq:EHartree} (line 5), just for the vector potential,
\begin{equation}\label{eq:def-bEM}
\begin{aligned}
    &\bEM(v,\AA) = \EM(v,\AA) - \frac{1}{2}\|\AA\|^2_{Y^*} \\
    &= \inf_{\aa\in Y^*} \left\{ E(v,\aa) + \frac{1}{2}\|\aa-\AA\|_{Y^*}^2 \right\} - \frac{1}{2}\|\AA\|^2_{Y^*}.
\end{aligned}
\end{equation}
This functional is now jointly concave in $v$ and $\AA$, yet we do \emph{not} have the same effect of auto-regularization as for $\EHartree$ since $E(v,\aa)$ is not convex (nor concave) in $\aa$. For a regularized functional a further step is necessary.

By Legendre--Fenchel transform in just one or both arguments, one can construct three further functionals,
\begin{align}
    &\bFM(\rho,\AA) = \sup_{v\in X^*}\{ \bEM(v,\AA)-\langle v,\rho \rangle \}, \\
    &\beM(v,\jp) = \sup_{\AA\in Y^*}\{ \bEM(v,\AA)-\chi\langle \AA,\jp \rangle \}, \\
    &\bfM(\rho,\jp) = \sup_{\AA\in Y^*}\{ \bFM(\rho,\AA)-\chi\langle \AA,\jp \rangle \}.
\end{align}
In this step we introduced a new magnetic coupling constant $\chi$. Now it turns out that the last two functionals, $\beM(v,\jp)$ and $\bfM(\rho,\jp)$ can be given in the form of Lasry--Lions regularizations~\cite{Lasry-Lions1986}, a form of double MY regularization, that has the formula
\begin{equation}\label{eq:LL}
    f_{\eps,\delta} (x) = \sup_{z\in X}\inf_{y\in X}\left\{ f(y) + \frac{1}{2\eps}\|x-y\|_{X}^2 - \frac{1}{2\delta}\|x-z\|_{X}^2 \right\}.
\end{equation}
By a theorem due to \citet{Attouch-Aze1993} they both become Fréchet differentiable if $\chi>1$. This differentiability allows for a unique mapping from paramagnetic currents to vector potentials $\AA$, either for given $v$ or $\rho$, and thus gives a form of Hohenberg--Kohn theorem in the context of Maxwell--Schrödinger DFT. The details of this procedure can be found in the original publication by \citet{tellgren2018-MDFT}.

If we remember that the Coulomb potential arises from a QED treatment of the longitudinal component of the electric field in Coulomb gauge~\cite{greiner-reinhardt2013}, then the relation between the Hartree approximation (with only mean-field interaction) and standard DFT is the same as between Maxwell--Schrödinger DFT and a \emph{full} (non-relativistic) QED treatment in terms of a functional theory. The development of such a theory, called QEDFT, is an ongoing program~\cite{ruggenthaler2014quantum,ruggenthaler-QEDFT-arxiv,Lu2024,Bakkestuen2025,Bakkestuen2025-Dicke} with important implications for fields like polaritonic chemistry~\cite{Ruggenthaler2023}, and it might greatly benefit from MY regularization on its theoretical side.

\section{Application of MY-based Kohn--Sham inversion to periodic settings}
\label{sec:MY-periodic}

Crucial for the density--potential inversion described in Section~\ref{sec:inversion}
is the computation of the proximal point $\prox_{\eps F^\lambda}(x)$.
For the full universal density functional, including interactions,
this is a daunting task.
Considering the inversion for the non-interacting functional $F^0(\rho)$
is considerably easier and the resulting method is called ``inverse KS'' since it relates to the KS system.
For a periodic setting a KS inversion algorithm based
on MY regularization has recently been implemented
and analyzed leading to first error bounds~\cite{Herbst2025}.

A key advantage of the periodic setting for exploration is that natural
choices for the density and potential spaces, $X$ and $X^*$,
are the periodic Sobolev spaces $H_\mathrm{per}^s(\Omega)$ with $\Omega \subset \R^3$ denoting the periodically repeated unit cell~\cite[Sec.~2]{Cances2012}.
These spaces are Hilbert spaces (thus fulfill Assumption~\ref{ass:HS})
and are easily characterized mathematically using Fourier modes
$e_{\GG} = e^{\rmi\GG\cdot\rr} / \sqrt{|\Omega|}$, namely by
\begin{equation}\label{eq:Hs_per}
    H_\mathrm{per}^s(\Omega) = \left\{
        u = \sum_{\GG} \hat{u}_{\GG} e_{\GG} \middle|
        \hat{u}_{\GG}^*=\hat{u}_{-\GG},
        \norm{u}_{H_\mathrm{per}^{s}} < \infty
    \right\}
\end{equation}
with
\begin{equation}\label{eq:Hs_per-norm}
    \norm{u}_{H_\mathrm{per}^{s}}^2= \sum_{\GG} (1 + |\GG|^{2})^s |\hat{u}_{\GG}|^2,
\end{equation}
where the sum is over the usual reciprocal lattice vectors $\GG$.
This characterization is very much in line with the usual discretization
schemes based on plane waves, where essentially one just selects a subspace of $H_\mathrm{per}^s(\Omega)$ by limiting the largest $|\GG|$.
Note that the space with $s=+1$ is different to the homogeneous Sobolev space of Eq.~\eqref{eq:hom-Sobolev}, apart from consisting of periodic functions, since the $1+|\GG|^{2}$ (instead of just $|\GG|^{2}$) in the definition of the norm in Eq.~\eqref{eq:Hs_per-norm} means the uniform $\GG=0$ component is included as well.

As a result of the above definition of the norm in terms of a Fourier expansion, key quantities of the MY program, such as the duality map $J$, can be obtained
in computable, closed-form expressions.
For example, with the particular choice $X = H_\mathrm{per}^{-1}(\Omega)$ for densities and $X^* = H_\mathrm{per}^{1}(\Omega)$ for potentials
one obtains the duality map in the Fourier representation as~\cite{Herbst2025}
\newcommand{\DS}{H_\mathrm{per}^{-1}}
\newcommand{\VS}{H_\mathrm{per}^{1}}
\begin{equation}\label{eq:Jfourier}
   J(\rho) = \sum_{\GG} \frac{\hat{\rho}_{\GG} e_{\GG}}{1 + \abs{\GG}^2}.
\end{equation}
Noticing that $1 / (\mu^2 + \abs{\GG}^2)$ is the Fourier representation of the Yukawa potential
$\Phi^\mu(\rr) = \exp{(-\mu \abs{\rr})}/ (4\pi \abs{\rr})$ then admits the
closed-form formulation as a convolution,
\begin{equation}\label{eq:Jrealsp}
    J(\rho)(\rr) = (\Phi^{\mu=1} \ast \rho )(\rr) = \int_{\R^3} \frac{e^{-|\rr - \rr'|} \rho(\rr')}{4\pi|\rr - \rr'|} \rmd\rr',
\end{equation}
which is efficiently computable in a plane-wave discretization
using Eq.~\eqref{eq:Jfourier} and fast-Fourier transforms.
We remark that the validity of Assumption~\ref{ass:DFT}
must still explicitly be demonstrated with this choice of spaces, although the periodic setting naturally guarantees the existence of a ground state for every choice of potential $v\in H_\mathrm{per}^1(\Omega)$.

Targeting KS inversion in this setting we (as usual) assume that we are given a simultaneously interacting and non-interacting $v$-representable ground-state density $\rhogs$, meaning that there exists
a KS potential $v_\rms \in -\partial F^0(\rhogs)$ and a potential $v \in -\partial F^1(\rhogs)$.
While $F^0(\rho)$ could be directly subjected to the MY program,
results seem to generally improve if a maximal amount of known information about the problem
is explicitly employed~\cite{Zhao1994,Shi2021}.
We therefore choose a `guiding functional'
\begin{equation} \label{eq:invguide}
    \Fguide(\rho) = F^0(\rho) + \EH(\rho) + \langle v,\rho \rangle,
\end{equation}
which additionally includes the Hartree mean-field energy and the external potential.
Noting that $\EH(\rho)$ in the periodic setting is modified to include a compensating background,
that is $\EH(\rho) = \sum_{\GG \neq 0} \abs{\hat{\rho}_{\GG}}^2/\abs{\GG}^2$,
one shows $\Fguide(\rho)$ to remain convex~\cite{Herbst2025}.
Lower semicontinuity of $F^0(\rho)$ and thus the entire functional $\Fguide(\rho)$
in the $\DS$ topology remains to be shown.
Similarly, we remark that the ideal choice of guiding functional
is currently also an open research question.

Using property~\ref{item:diff-f_eps} of MY regularization in Section~\ref{sec:MY} then yields
\begin{equation}
    \eps^{-1} J (\rhogseps - \rhogs) \in
    -\partial \Fguide(\rhogseps),
\end{equation}
where we introduced the shorthand $\rhogseps = \prox_{\eps\Fguide}(\rhogs)$.
We remark that our assumptions on $\rhogs$ allow the definition of an exchange--correlation potential 
$\vxc = v_{\rm s} - v - \vH(\rhogs) \in -\partial \Fguide(\rhogs)$
that we obtain in the limit $\eps\to 0$ (just like in Section~\ref{sec:prox-ZMP})
\begin{equation}\label{eq:MYpot}
    \vxc = \lim_{\eps\to 0} \vxceps
    \quad \text{with} \quad \vxceps = \eps^{-1} J(\rhogseps -\rhogs ).
\end{equation}
To compute this numerically we require two ingredients:
\begin{enumerate}[(i)]
    \item\label{ingredient:prox} an approach to compute the proximal points $\rhogseps$,
        which we will discuss next, and,
    \item\label{ingredient:eps} a suitable sequence of decreasing values for $\eps$
enabling an extrapolation of the $\eps\to0$ limit.
    Despite progress in the development of analytical error bounds
        for MY-based KS inversion~(MYiKS), the choice of such an $\varepsilon$-sequence remains empirical to date.
        An explicit convergence study is thus still required as we will outline below.
\end{enumerate}

One approach to achieve \ref{ingredient:prox} is to follow the fixed-point iteration suggested in Eq.~\eqref{eq:prox-iter}.
However, in our previous work~\cite{Herbst2025}, we directly
solve the minimization problem encoded in the proximal-point computation,
i.e., we minimize
\begin{equation}
    \Fguide(\rho) + \frac{1}{2\eps} \| \rho - \rhogs \|_{\DS}^2
\end{equation} over all $\rho\in D$,
which automatically excludes non-physical densities (e.g., with infinite kinetic energies) from the minimization.
Notably, for the numerical approximation of the
supremum in Eq.~\eqref{eq:LegendreFenchel} for $F^0(\rho)$ we employ the usual KS parametrization of the density
in terms of orthonormal orbitals. This parametrization reads $\Phi=(\phi_1,\ldots,\phi_N)$
with $\langle \phi_i, \phi_j \rangle = \delta_{ij}$
and occupations $\bm{f} = (f_1,\ldots,f_N)$ and gives
\begin{equation} \label{eq:dens}
    \rho_\Phi(\rr) = \sum_{i} f_i |\phi_i(\rr)|^2.
\end{equation}
We can then write the proximal-point computation as the minimization of
\begin{equation}\label{eq:Epractical}
\begin{aligned}
    \mathcal{E}(\Phi, \bm{f})
    =&\, \sum_{i} f_i \| \nabla \phi_i\|^2_{L^2}  + E_\mathrm{H}(\rho_\Phi) \\
    &+ \langle v, \rho_\Phi \rangle + \frac{1}{2\eps} \| \rho_\Phi - \rho_\mathrm{gs} \|_{\DS}^2
\end{aligned}
\end{equation}
with respect to the orbitals $\phi_i$ and occupations $f_i$.
In these expressions we 
do not make explicit any details about spin or Brillouin zone sampling,
i.e., the choice of the $\kk$-point grid to model periodic supercells~\cite[Ch.~2.8]{Lin2019}.
In that sense one could think of the index $i$ to run over all occupied orbitals
of all spins and all reducible $\kk$-points.

We remark that this energy expression Eq.~\eqref{eq:Epractical} is closely related
to the usual KS-DFT energy expression, where essentially the exchange--correlation energy
term is just replaced by the penalty term of the regularization.
Standard direct minimization techniques from KS-DFT are therefore immediately applicable%
~\cite{Payne1992,Kresse1996,Dai2017}.
A first implementation of this approach in DFTK~\cite{Herbst2025,DFTKpaper}
is capable of performing proximal-point computations for 
non-trivial 3D solid-state systems, such as silicon, gallium arsenide
or potassium chloride, thus providing a practical MYiKS procedure.
Right now the treatment is still restricted to insulating solids,
since the algorithm currently makes the assumption that all $f_i = 1$ in Eq.~\eqref{eq:dens} and
respectively only minimizes over the $N$ (spin) orbitals in Eq.~\eqref{eq:Epractical}.
A generalization to $f_i \neq 1$, required for the treatment of metals,
is possible taking inspiration from the KS-DFT literature~\cite{Kresse1996,Marzari1997,Freysoldt2009,Gonze2024}.

While the current practical MYiKS scheme is still far from optimized
in terms of employed numerical algorithms and implementation,
we want to provide a high-level idea of their computational cost.
Similar to standard KS-DFT, the main computational bottlenecks of MYiKS
are the matrix-vector products involving the gradient of $\mathcal{E}$,
which is an operator similar to the usual KS Hamiltonian.
For plane-wave basis sets a gradient-vector product
scales as $O(N_b \log N_b)$, where $N_b$ is the number of plane waves.
One iteration step in the minimization of Eq.~\eqref{eq:Epractical} typically requires
one gradient-vector product.
With current algorithms a few tens (for large $\eps$) to a few hundred (for small $\eps$)
of minimization steps are required to compute one proximal density.
The advantage of the MYiKS approach is therefore that no explicit diagonalization
of a KS-like Hamiltonian is required---in contrast to for example the ZMP~\cite{Zhao1994} or the Wu--Yang~\cite{Wu2003} methods.
However, 
the disadvantage of current MYiKS techniques are (i) the sometimes
large number of minimization steps per proximal point computation
and (ii) the need to estimate the limit $\eps\to0$, i.e., the need to perform 
multiple proximal point computations for different $\eps$ values.
We now describe some ideas to overcome these challenges.
%

Crucial to reduce the number of minimization steps is an appropriate preconditioning
of the minimization of $\mathcal{E}$. 
For the usual KS problems it is primarily the kinetic-energy term
with its $|\GG|^2$-dependence that leads to
ill-conditioned problems in large plane-wave basis sets.
Appropriate preconditioners have been developed to effectively tame this source
of ill-conditioning~\cite{Teter1989,Kresse1996} for KS DFT problems.
For the MY proximal-point computation we currently rely on the
same kinetic-energy preconditioners.
For large $\eps$ values this is indeed sufficient.
However, for small $\eps$ the penalty term
$\frac{1}{2\eps} \| \rho_\Phi - \rho_\mathrm{gs} \|_{\DS}^2$
can dominate, such that kinetic-energy preconditioning alone
is no longer appropriate. 
This explains why sometimes our minimization procedure requires hundreds of steps.
Refined preconditioning techniques for the MY setting 
therefore need to be developed.
These should be $\eps$-dependent
and take both the kinetic-energy and the penalty term
into account.

\begin{figure}
    \includegraphics[width=0.5\textwidth]{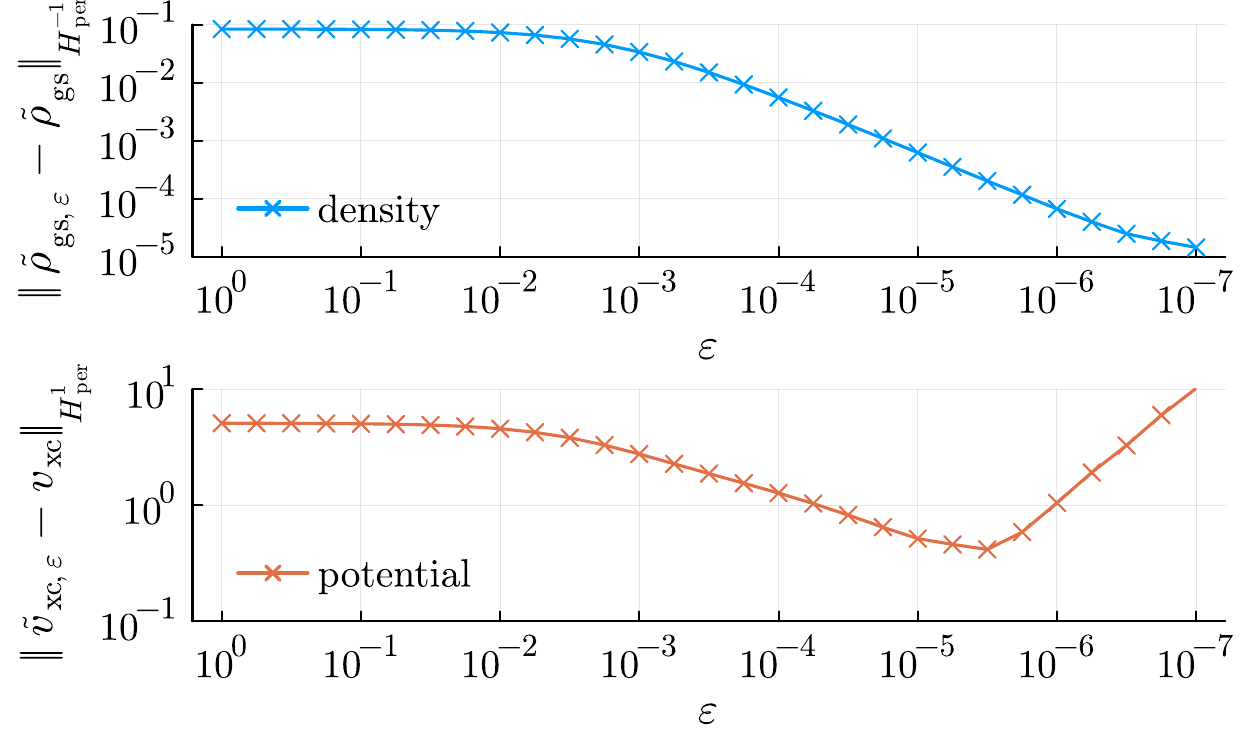}
    \vspace{-2.3em}
    \caption{Convergence of $\tilde{\rho}_{\text{gs},\eps}$ and $\tvxceps$
        as a function of $\eps$ for the bulk silicon test
        case and a basis truncation
        of $E_\text{cut} = 30$, leading to an error
        $\delta \rho = 2.7 \times 10^{-5}$.
        We refer the reader to the original reference~\cite{Herbst2025}
        for details of the computational setup and numerical parameters.
    }
    \label{fig:epsilonconvergence}
\end{figure}
An additional complication in practical inverse KS calculations
is that the input density $\rhogs$ itself is usually obtained from an inexact procedure,
such as an experiment or a computation from a higher-level theory.
As a result, we typically do not have access to the exact $\rhogs$,
but only to a perturbed version $\tilde{\rho}_\text{gs} = \rhogs + \delta \rho$.
Computing the proximal point for a given $\eps$ with this perturbed ground-state density
thus only provides us with an estimate $\tvxceps$ instead of the correct $\vxceps$.
Previous analysis~\cite{Herbst2025} lets us deduce the bound
\begin{equation}\label{eqn:errorestimate}
\begin{aligned}
    \norm{\vxc - \tvxceps}_{\VS}
    &\leq \norm{\vxc - \vxceps}_{\VS}
    + \norm{\vxceps - \tvxceps}_{\VS} \\
    &\leq \norm{\vxc - \vxceps}_{\VS}
    + \frac{C}{\eps} \norm{\delta \rho}_{\DS}
\end{aligned}
\end{equation}
for the error between the numerically computed $\tvxceps$ and the desired true $\vxc$. 
We notice two error contributions.
The last term is due to the inherent error $\delta \rho$ in the available
ground-state density data and increases with decreasing $\eps$. 
We observe (see Figure~\ref{fig:epsilonconvergence}) that
the potential $\tvxceps$ initially converges towards the true $\vxc$,
but then the distance increases again: 
there is an optimal $\eps$ value
which yields the most accurate estimate of the inverted potential.
We remark that on top of the effects of a noisy input density outlined
in Eq.~\eqref{eqn:errorestimate} and
which we studied in Ref.~\onlinecite{Herbst2025},
there is an additional contribution to the total error that we have not made explicit here.
This is the error due to the proximal density
being only converged up to a certain finite tolerance in practical iterative schemes
for minimizing $\mathcal{E}$.
Including this term consistently in the analysis is work in progress.

The bound in Eq.~\eqref{eqn:errorestimate} provides a qualitative explanation
of the observed behavior as $\eps\to0$. However,
it is neither sharp nor does it fully match the practical setting of periodic calculations,
such that it is not yet sufficient to suggest an optimal $\eps$ value \textit{a priori},
let alone an entire $\eps$ sequence for potential extrapolation.
For example, while Eq.~\eqref{eqn:errorestimate} lets us observe that $\eps$ should not be taken smaller than the error $\delta \rho$, this error is usually not known in practice.
Furthermore, the use of a pseudopotential approximation is required for efficient plane-wave calculations.
This implies the addition of a non-local (Kleinman--Bylander) term~\cite[Ch.~7]{DftBook}
to the energy functional Eq.~\eqref{eq:Epractical},
which is outside of the scope of the current mathematical framework~\cite{Herbst2025}.

While the current analysis thus already provides first insights
into the optimal choice of an $\eps$ sequence (ingredient \ref{ingredient:eps}),
it still requires considerable refinement until an \textit{a priori} mathematically
guided selection of such a sequence is feasible.
To date, an explicit numerical study of the $\eps\to0$ limit,
e.g., based on a logarithmically spaced $\eps$ grid,
is thus generally required to empirically identify the optimal $\eps$.
However, the progress towards first analytical error bounds such as Eq.~\eqref{eqn:errorestimate}
as well as the rich literature on \textit{a priori} and \textit{a posterior} error analysis
for the closely related problem of periodic KS-DFT~\cite{Cances2010,Cances2012,Cances2018,scferror,Cances2020,Dusson2023,Bordignon2024,Cances2024}
make it hopeful that a more refined and comprehensive analysis is feasible.
A tantalizing outlook would be sharper error bounds and thus a better estimation of an optimal $\eps$,
or the combination of convergence properties of $\tvxceps$
with extrapolation techniques to accelerate convergence.

Beyond refinements of the inversion algorithm itself,
a number of possible alternative choices in the mathematical setup have neither been fully explored.
This includes a different choice than $\DS(\Omega)$ for the density space $X$ and $\VS(\Omega)$ for the potential space $X^*$.
For example, the inverse choice
($\DS(\Omega)$ for the potential and $\VS(\Omega)$ for the density) would in principle work as well
and exactly fits to recent results about $v$-representability
on periodic one-dimensional domains~\cite{Sutter2024,CarvalhoCorso2025}.
Yet, it has to be noted that this will break the self-adjointness requirement from Assumption~\ref{ass:DFT} in more than one dimension~\cite{HERCZYNSKI1989-KLMN}.
We believe that systematically exploring the question of function spaces
is a promising endeavor to precisely target the inversion procedure
towards reproducing the physically relevant features of the potential:
as visible from Eq.~\eqref{eq:Epractical}, the function space choice is directly reflected
in the norms employed during the proximal point computation.
From a physical point of view, the regions close to the nucleus are generally less relevant,
such that an ideal function space $X$ should feature a norm that puts less weight
on the core and larger weight on the valence regions.

In closing, we remark that the exceptionally close agreement between the
plane-wave discretization basis and the Fourier characterization
of the periodic Sobolev spaces make the periodic setting
ideally suited for further joint numerical and analytical study of MY-regularized DFT.
In combination with a flexible tool such as DFTK~\cite{DFTKpaper},
mathematical ideas can be seamlessly implemented and tested both on toy problems
as well as realistic materials systems.
Considering specifically the
similarity of MY-based KS inversion
to other inversion approaches (such as the ZMP method), see Section~\ref{sec:inversion},
as well as the critical lack of mathematical analysis for practical inverse KS,
this provides an accessible
framework to close this important gap.


\section{Outlook}
\label{sec:outlook}

This work aimed at showing how, since its introduction into the field of DFT, the technique of MY regularization grew ever more useful and versatile. At this point, it is interesting to ask, which are the most promising routes to follow and what more can be expected. Here, we collect a number of ideas and possible research directions that we deem especially interesting and worthwhile to investigate. This will at the same time serve as a summary of the presented concepts relating to MY regularization applied to DFT.

\begin{itemize}[leftmargin=*]
    \item In Section~\ref{sec:autoreg} it was demonstrated how in certain mean-field approximations a choice of spaces for densities and potentials that is adapted such that $\|\cdot\|_{X^*}^2$ yields the field energy automatically leads to MY regularization. A very similar choice of spaces was made in the ZMP method for density--potential inversion (Section~\ref{sec:inversion}) and in the numerical discussion of the periodic setting (Section~\ref{sec:MY-periodic}). In general, we believe that fitting the mathematical setting to physical principles, while taking all the functional-analytic considerations of Section~\ref{sec:formalDFT} into account, is extremely useful as a guiding principle. The original Lieb setting~\cite{Lieb1983} with $X=L^1\cap L^3$ in some sense also had this in mind via the estimate of the $L^3$-norm of the density by the kinetic energy, but this is just comparative, while an exact relation to the field energy is possible. Equation~\eqref{eq:J-Poisson} demonstrated how this leads to an equivalence between the duality map and the electrostatic Poisson equation. It is thus possible to \emph{encode important physical relations fully into the geometry} of the density and potential spaces. (Note that such an encoding of physical laws into geometry is also present quite generally in gauge theories through the introduction of fibre bundles.) This basic idea will be pursued further for a new formulation of QEDFT in a forthcoming publication, where the adaption of the norm to the field energy means that all of Maxwell's equations will be intimately linked to the theory. This then gives rise to the exact macroscopic Maxwell formulation of non-relativistic QED and has important consequences for renormalization.

    \item The spaces of Section~\ref{sec:autoreg} just mentioned also happen to be Hilbert spaces, thus fulfilling our Assumption~\ref{ass:HS}. This means that additionally to the Banach-space structure several extremely beneficial properties hold true, like the linearity of the duality map and firm nonexpansiveness of the proximal map. Up to this point, this was not used to full capability. It is for example possible that the Hilbert-space setting can help to generalize the proof of convergence of the regularized KS scheme from Section~\ref{sec:epsKS-convergence} to infinite dimensions.

    \item A generalization in the opposite direction is also possible. This means considering Banach spaces and adapting the MY regularization itself to what is called the `modulus of convexity' of the respective space. This is possible by replacing the $(\cdot)^2$ in Eq.~\eqref{eq:MY} by $(\cdot)^p$, $p>1$, or right away by a Young function $\phi(\cdot)$. One then also needs to switch to a corresponding generalized duality map according to Eq.~\eqref{eq:J-from-subdiff}. While this direction was explored mathematically~\cite{bacak-kohlenbach-2018,Bacho2023,Penz2025-pMY}, it was not really implemented into the study of DFT yet.
    
    \item The original motivation for devising the regularized KS scheme was to be able to prove its convergence. Yet, clearly, a guaranteed convergence is also beneficial for practical applications, so one could try and develop functional approximations $\vxceps(\rho)$ that are already adapted to the regularized KS iteration of Section~\ref{sec:epsKS}. A small step in this direction was presented in Section~\ref{sec:transform-xc} that demonstrated a way how to transform available exchange--correlation potentials such that they can be applied in the regularized KS scheme. But this must be seen just as a first approach to the question. Instead, a construction of such approximations from first principles could also be tried. On the other hand, the practical implications of switching to the regularized form of the KS method have not really been explored yet. 
    
    \item Speaking of convergence, to date we do not have a convergence proof for the proximal-point iteration (Section~\ref{sec:prox-ZMP}) that is also the basis for the inverse KS method applied to periodic settings (Section~\ref{sec:MY-periodic}). We can maybe expect it in the form of a fixed-point proof that is typical for Banach-space structures. We also note at this point that the presented method can easily be adapted to other DFT settings, like spin DFT, current DFT, or QEDFT.

    \item As mentioned, the homogeneous Sobolev spaces introduced in Section~\ref{sec:autoreg} and their duals might have a central role in future reformulations of (regularized) DFT, but they are not that well studied in functional analysis, especially not on the dual side. What is known is that they are Hilbert spaces and thus Assumption~\ref{ass:HS} is secured. We noted in Section~\ref{sec:MY-mixing} that they are already used in some practical schemes to enhance the Kohn--Sham method. Further functional theoretic insights on these spaces would thus be a valuable contribution to ongoing work. In the course of such study, the important questions of self-adjointness and boundedness below of the Hamiltonian and, in particular, the lower semicontinuity of the ensemble constrained-search density functional must also be clarified in order to have the correct setting for MY-regularized DFT according to Assumption~\ref{ass:DFT}.
    In Section~\ref{sec:formalDFT} it was noted that possible strategies to have a lower semicontinuous $\FDM^\lambda(\rho)$ consist in choosing a bounded domain $\Omega$ (where it is possible to add periodic boundary condition) or to include an offset potential like $\vo(\rr)=|\rr|^2$. Both options are currently investigated.

    \item The whole regularized KS framework in Section~\ref{sec:epsKS} used the difference between the interacting functional $F^1_\eps$ and the non-interacting $F^0_\eps$, as typically done in KS-DFT. Yet, it was noted in Section~\ref{sec:MY-periodic} that switching to a different guiding functional that already includes \emph{a priori} information might be beneficial at least numerically. In a way we did this already in Section~\ref{sec:formalDFT} through the inclusion of the offset potential $\vo$ that would include information about the external potential. But other, more complicated terms, like the Hartree mean-field energy $\EH(\rho)$ could be included as well before regularization, thus choosing $(F^0 + \EH)_\eps$ as a regularized guiding functional. One only has to make sure that the functional is still convex and lower semicontinuous before regularization, in order to stay within Assumption~\ref{ass:DFT}. Further studies about these questions are already underway.

\item In the periodic setting, the computation of proximal points, respectively the inversion of Kohn--Sham densities based on MY techniques has recently become feasible for practical systems, see Section~\ref{sec:MY-periodic}. However, the currently considered algorithm still suffers from a number of restrictions: (i) it is only applicable to insulating systems,
    (ii) proper preconditioners have not yet been developed, and
    (iii) the choice of regularization parameter $\eps$
    requires careful manual tuning or an explicit convergence study.
    Overcoming these restrictions is crucial to broaden the scope
    of MY-based numerical experiments for mathematical research on DFT
    as well as for practical KS inversion computations.

\item While recasting KS inversion via MY regularization yields an \textit{a priori}
    orbital-free density-functional formulation,
    see Section~\ref{sec:inversion}, practical implementations of MY-based KS inversion (MYiKS)
    still rely on an explicit orbital parametrization
    to solve the proximal-point optimization problem Eq.~\eqref{eq:Epractical}.
    The same is true for many standard KS inversion algorithms~\cite{Zhao1994,Wu2003,Shi2021}.
    Consequently, it would be advantageous to develop a mathematical framework
    in which the regularization is expressed directly in terms of the orbitals
    to provide a closer match between MY-based theory and practical inversion schemes.
    We remark that such an orbital-based formulation is also more naturally suited
    to provide a fully consistent treatment of the nonlocal term present in most
    pseudopotential approximations~(see Section~\ref{sec:MY-periodic}).

\item As previously discussed in Section~\ref{sec:MY-periodic}, the close match
    between the mathematical framework and the numerical implementation in the
    periodic setting is remarkable. This enables joint analytical and numerical
    investigations, where mathematical developments can be guided and tested
    using numerical experiments, and suggested theoretical relationships can be
    directly exploited in the development of inversion algorithms with improved
    efficiency and reliability.
    Considering the many possibilities for
    formulating MY-regularized DFT (e.g., the considerable flexibility in
    selecting appropriate function spaces), we further expect numerical
    investigations to be decisive for suggesting which theoretical extensions
    are worthwhile to pursue. Moreover, we remark that since the
    MY-based Kohn--Sham inversion framework can be viewed as a unified framework
    encompassing a wide range of inversion schemes, see Section~\ref{sec:inversion},
    we expect the resulting mathematical and algorithmic advances from such studies to be
    broadly applicable and to directly contribute to the development of reliable
    and computationally efficient inversion schemes.
\end{itemize}

\newcommand{\nocontentsline}[3]{}
\let\origaddcontentsline\addcontentsline
\newcommand{\stoptoc}{\let\addcontentsline\nocontentsline}
\newcommand{\resumetoc}{\let\addcontentsline\origaddcontentsline}
\stoptoc

\section*{Data availability}
No new data were created or analyzed in support of this research.

\begin{acknowledgments}
The authors thank Erik I.\ Tellgren, Mihály A.\ Csirik, Thiago Carvalho Corso, and Michael Ruggenthaler for valuable discussions.
MP acknowledges support from the German Research Foundation (Grant SCHI 1476/1-1).
MFH acknowledges support by the Swiss National Science Foundation (SNSF, Grant Nos.~221186 and 10002757) as well as the NCCR MARVEL, a National Centre of Competence in Research, funded by the SNSF (Grant No. 205602). 
AL and MP have received funding from the ERC-2021-STG under grant agreement No.~101041487 REGAL. AL and TH were funded by the Research Council of Norway through CoE Hylleraas Centre for Quantum Molecular Sciences Grant No.~262695.
\end{acknowledgments}

\bibliography{refs}
\end{document}